\pdfoutput=1
\documentclass[journal]{IEEEtran}
\usepackage{graphicx}
\usepackage{epstopdf}
\usepackage[cmex10]{amsmath}
\usepackage{cases}
\usepackage[tight,footnotesize]{subfigure}
\usepackage{amsthm}
\usepackage{cite}
\usepackage{amssymb}
\usepackage{caption}
\usepackage{algorithm}
\usepackage{algorithmic}
\usepackage{multirow}
\usepackage{amsmath}
\usepackage{xcolor}
\usepackage{CJK}
\usepackage{subeqnarray}
\usepackage{cases}
\usepackage{longtable}
\usepackage{multicol}
\usepackage{authblk}

\date{}   % ������ȥ�����ɵ�ʱ��
\ifCLASSINFOpdf
\else  % or other class option (dvipsone, ecified.
\fi
\begin{document}
\title{Intelligent Reflecting Surface Aided Mobile Edge Computing With Binary Offloading: Energy Minimization for IoT Devices}

\author{Yizhen Yang, Yi Gong, and Yik-Chung Wu
\thanks{
Yizhen Yang is with the Department of Electrical and Electronic Engineering, The University of Hong Kong, Pokfulam Road, Hong Kong, and also with the Department of Electrical and Electronic Engineering, Southern University of Science and Technology, Shenzhen 518055, China.

Yi Gong is with the University Key Laboratory of Advanced Wireless Communications of Guangdong Province, Department of Electrical and Electronic Engineering, Southern University of Science and Technology, Shenzhen 518055, China (e-mail: gongy@sustech.edu.cn).

Yik-Chung Wu is with the Department of Electrical and Electronic Engineering, The University of Hong Kong, Pokfulam Road, Hong Kong (email: ycwu@eee.hku.hk).

Yi Gong and Yik-Chung Wu are the corresponding authors.

Copyright (c) 20xx IEEE. Personal use of this material is permitted. However, permission to use this material for any other purposes must be obtained from the IEEE by sending a request to pubs-permissions@ieee.org.
}}

%\thanks{The authors are with the Ministry of Education Key Lab for Intelligent Networks and Network Security, School of Electronic and Information Engineering, Xi'an Jiaotong University, Xi'an 710049, China (e-mail: zongzeleo@gmail.com;pcmu@mail.xjtu.edu.cn; zongmian.li@gmail.com; antaeusas@stu.xjtu.edu.cn)}% <-this % stops a space
%\thanks{Manuscript received XXX, XX, 2016; revised XXX, XX, 2016.}


\maketitle

\begin{abstract}
Mobile edge computing (MEC) is envisioned as a promising technique to support computation-intensive and time-critical applications in future Internet of Things (IoT) era. However, the uplink transmission performance will be highly impacted by the hostile wireless channel, the low bandwidth, and the low transmission power of IoT devices. Recently, intelligent reflecting surface (IRS) has drawn much attention because of its capability to control the wireless environments so as to enhance the spectrum and energy efficiencies of wireless communications. In this paper, we consider an IRS-aided multidevice MEC system where each IoT device follows the binary offloading policy, i.e., a task has to be computed as a whole either locally or remotely at the edge server. We aim to minimize the total energy consumption of devices by jointly optimizing the binary offloading modes, the CPU frequencies, the offloading powers, the offloading times and the IRS phase shifts for all devices. Two algorithms, which are greedy-based and penalty-based, are proposed to solve the challenging nonconvex and discontinuous problem. It is found that the penalty-based method has only linear complexity with respect to the number of devices, but it performs close to the greedy-based method with cubic complexity with respect to number of devices. Furthermore, binary offloading via IRS indeed saves more energy compared to the case without IRS.

\end{abstract}


% \fi
\IEEEpeerreviewmaketitle

\section{Introduction}
\IEEEPARstart{I}{n} traditional mobile cloud computing systems, the data of mobile devices would be sent to the cloud server in the core network for further computing \cite{1}, \cite{2}. However, this scheme cannot fit the future Internet of Things (IoT) era due to the explosively increasing amount of data generated by massive number of IoT wireless devices and the time-critical requirements of new applications such as industrial monitoring, disaster early warning and healthcare \cite{31, 32, 33}. Recently, a new computing paradigm called \textit{mobile edge computing} (MEC) has emerged and drawn a lot of attention from both academia and industry \cite{3, 15, 16}. It pushes the computing capability from the core network to the network edge. In this context, IoT devices can offload their intensive computation to the nearby edge server, which is co-deployed with the access point (AP). By offloading computation to the network edge instead of the cloud, IoT devices can be served with ultra-low latency and core link congestion can be mitigated significantly \cite{6}.
\par There are two models in computation offloading: partial offloading and binary offloading \cite{4}. Specifically, partial offloading allows a task to be partitioned into two parts. One part is executed locally and the other is transmitted to the server for computation. On the other hand, binary offloading requires a task as a whole to be computed either locally or remotely at the server. An important aspect in computation offloading is the application model/type since it determines whether partial offloading is applicable \cite{6}. In practice, binary offloading is easier to implement and suitable for the tasks that cannot be partitioned, while partial offloading is favorable for data that are bit-wise independent and can be arbitrarily divided into different groups \cite{4}, \cite{22}.
\par For computation offloading in MEC systems, minimization of execution delay and minimization of energy consumption are two reasonable goals. For example, \cite{7} and \cite{8} aim at minimizing the execution delay by making a binary offloading decision under single-user context. For partial offloading mode, \cite{18} jointly optimizes the offloading ratio, transmission power, and CPU frequency to minimize the execution latency, and \cite{21} investigates a latency minimization problem in a multi-user MEC system with joint communication and computation resource allocation. On the other hand, to minimize the total energy consumption of users, \cite{17} presents a time-slotted signaling structure in a multi-user binary offloading MEC system where users have different latency constraints and \cite{19}, \cite{23} minimize the weighted sum of terminal energy consumption by jointly optimizing the offloading ratio and radio resource allocation under partial offloading context.
\par While offloading would reduce computation energy from the IoT device perspective, the offloading operation itself might consume a lot of energy if the device is far away from the AP or there is a heavy blockage between the device and the AP. Especially, because of the stringent size constraint and the consideration of production cost, an IoT device usually carries a small battery, making it sensitive to energy consumption. The recent emerging intelligent reflecting surface (IRS) would be a promising technology to enhance the energy and spectrum efficiencies in this case \cite{5}, \cite{9}. Specifically, an IRS consists of a large number of passive reflecting elements (e.g., positive-intrinsic-negative (PIN) diodes), and each of them is capable to reflect the incident signal with an adjustable phase shift. This reconfigures the signal propagation environment, thus significantly enhancing the quality of wireless communication. From this point of view, IRS is remarkably different from other prior communication techniques, which improve the quality of wireless communication via optimization at the transmitter or receiver. Furthermore, an IRS does not need any active module but only reflects signals passively, which incurs no transmission power consumption, and therefore is energy-economical compared to other communication technologies.
\par Extensive research efforts have been made to demonstrate the benefits of employing IRS in various wireless systems. For example, \cite{10} presents an IRS-based signal hot spot for improving the communication quality near the IRS. Further, \cite{11} proposes a joint design of the AP beamforming and the IRS phase shifts to reduce the downlink transmission power compared to the scenario without IRS. Moreover, \cite{12} and \cite{24} consider IRS-aided multi-user MISO communication systems and aim at energy efficiency maximization and weighted sum-rate maximization, respectively. Besides, IRS has been deployed in orthogonal frequency division multiplexing (OFDM) wireless systems \cite{13}, \cite{25}, integrated with simultaneous wireless information and power transfer (SWIPT) technique \cite{26}, \cite{27} and applied in edge caching \cite{34} to remarkably improve the system performance. These existing research works demonstrate the benefits of IRS to wireless communications.
\par While IRS is widely employed in wireless communication, the integration of IRS and computation offloading has seldom been considered. In fact, IRS can be regarded as a kind of communication resource due to its capability to improve the quality of wireless channels. Thus, it can be applied in any wireless communication system to improve the system performance, including MEC system. For example, in an industrial monitoring system, the data collected by the monitoring devices will be transmitted to an edge server for further processing. In order to prolong the lifetimes of devices, IRS can be used under this scenario to improve the wireless environment so that the energy consumption of devices can be significantly reduced. Besides, in 5G or upcoming 6G cellular network, a lot of new applications will be implemented, such as autonomous vehicles, virtual reality and so on. In an autonomous driving system, a huge amount of data will be generated and transmitted to the edge server for processing. Under this scenario, IRS can be used to aid the wireless transmission between the autonomous vehicle and the edge server, so that the system latency can be reduced.
\par Recent works \cite{28} and \cite{29} consider employing IRS to aid MEC systems to minimize the latency and enhance the system sum computational bits, respectively. However, both \cite{28} and \cite{29} operate the MEC system in partial offloading mode. In contrast, there is a lack of studies in the design of IRS-aided binary offloading MEC system. In practice, not all computational tasks can be partitioned and binary offloading is easier to implement \cite{6}, so it is necessary to explore the benefits and effects of IRS in binary offloading MEC system.
\par In this paper, we consider an IRS-aided multiuser MEC system where each user is an IoT device with capacity-limited battery and follows the binary offloading policy. Specifically, in the considered system, within a certain time duration, each device has a computational task to be completed either locally or by offloading it to the edge server with the aid of IRS. To avoid inter-device interference, each offloading device would occupy a distinct time slot. An optimization problem is formulated to minimize the total energy consumption of IoT devices by determining the binary offloading modes, the CPU frequencies, the offloading powers, the offloading times, and the IRS phase shifts for all devices. The main difference from traditional MEC offloading is that the IRS phase shifts have to be jointly determined with the offloading / local computing decision. Due to the binary offloading variables, this optimization problem is non-convex and with discontinuities in the objective function, making it challenging to be solved. Furthermore, due to the constant modulus constraint of the IRS phase shift, the optimization problem is more complicated than that of traditional MEC offloading design.
\par To overcome the challenges, we first propose a greedy-based algorithm. Recognizing that the binary offloading decision is the most difficult part of the problem, this algorithm adds offloading devices one by one in a greedy sense. More specifically, under a particular offloading decision, we need to optimize the offloading powers, the offloading times, and the IRS phase shifts for the devices who offload data and optimize the CPU frequencies for the devices who compute locally. Once we optimize these parameters, the energy consumption of devices can be determined. Next, we look for a device that leads to a maximum reduction of energy consumption by switching it from computing locally to offloading. The process is repeated until we cannot find a device that leads to smaller energy consumption than the current solution. While the greedy algorithm is intuitive, this algorithm has an overall cubic computational complexity with respect to the number of devices, which is undesirable for systems with a large number of devices. To this end, a penalty-based algorithm is proposed. In contrast to the greedy-based algorithm, the penalty-based algorithm decomposes the original problem into many parallel subproblems so that the offloading decision for each device is determined in parallel. Thanks to this parallelization, the penalty algorithm has only linear computational complexity with respect to the number of devices, which is a huge advantage over the greedy-based algorithm in systems with a large number of devices.
\begin{figure}[htbp]
\centering
\includegraphics[width=3.5in]{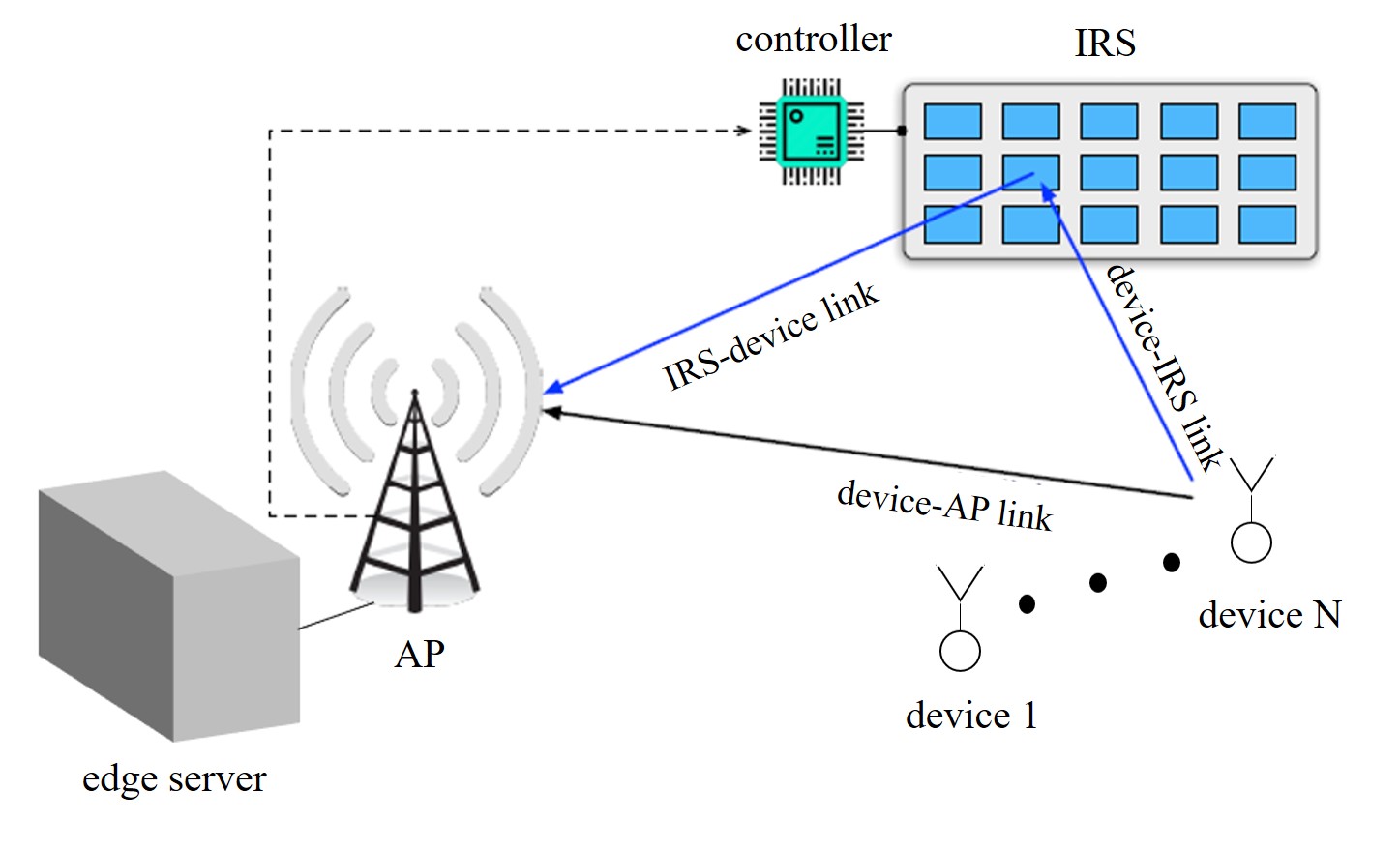}
\caption{An IRS-aided multi-device MEC system}
\end{figure}
\par Numerical results are provided to demonstrate the effectiveness of the two algorithms. It is found that the optimized binary offloading decision leads to a significantly smaller device energy consumption than mandatory offloading or mandatory local computation. Furthermore, deployment of IRS further reduces the energy consumption compared to systems without IRS. Finally, it is found that the greedy-based algorithm performs slightly better than the penalty-based algorithm at the cost of a much longer run time when the number of devices is large. Thus, for systems with a small number of devices, the greedy-based algorithm might be preferred, while for systems with a large number of devices, the penalty-based algorithm shall be adopted. The contributions of this paper are summarized as follows:
\par 1) we propose IRS-aided binary offloading MEC system for the first time in order to significantly improve the performance of MEC system. An energy-minimization problem is formulated to minimize the total energy consumption of IoT devices by optimizing the binary offloading modes, the CPU frequencies, the offloading powers, the offloading times and the IRS phase shifts for all devices. This formulation does not exist in the open literature.
\par 2) To tackle this challenging problem, two algorithms, which are the greedy-based and the penalty-based, are proposed. Especially, the penalty-based algorithm solves this problem with only linear computational complexity with respect to the number of IoT devices.
\par 3) Numerical results are presented to demonstrate the energy reduction effect by employing IRS, the effectiveness and characteristics of two algorithms. The influences of IRS on offloading decisions are also investigated.
\par The rest of this paper is organized as follows. The system model is introduced in Section II and the problem formulation and transformation is given in Section III. In Section IV and Section V, two algorithms are proposed to solve the discontinuous nonconvex problem. Section VI discusses the extension to discrete IRS phase shifts. Simulation results are presented in Section VII. Finally, Section VIII concludes the paper.
\section{System Model}
\subsection{Network Model}
As shown in Fig. 1, we consider a multi-device MEC system where a single-antenna AP (co-deployed with the edge server) serves $N$ single-antenna IoT devices over a certain frequency band. An IRS with $M$ reflecting elements is deployed to aid this system. Outfitted with a smart controller which is connected with the AP through a specialized wireless control link for information exchange, the IRS is able to independently adjust the phase shift of each reflecting element.
\begin{table}[H]
\scriptsize
\centering
\setlength{\abovecaptionskip}{0pt}
\setlength{\belowcaptionskip}{10pt}
\caption{NOTATIONS}
\begin{tabular}{|c|c|}
               \hline
                \textbf{Notation}  & \textbf{Description} \\
               \hline
                $T$ &a certain time duration \\
               \hline
                $S_n$ &the data size of the computational task of device $n$ \\
               \hline
                $C_n$ &the number of CPU cycles for processing 1 bit of data for device $n$\\
               \hline
                $\varepsilon_n$ &the coefficient depending on the chip architecture of device $n$ \\
               \hline
                $f_n$ &the CPU computational frequency of device $n$ \\
               \hline
                $f_{n,max}$ &the maximum CPU frequency of device $n$ \\
               \hline
                $g_n$ &the channel from device $n$ to the server \\
               \hline
                $\textit{\textbf{h}}_n$ &the channel from device $n$ to the IRS \\
               \hline
                $\emph{\textbf{r}}$ &the channel from the IRS to the server \\
               \hline
                $\boldsymbol{\theta}_n$ &the phase shifts of all the reflecting elements dedicated to device $n$ \\
               \hline
                $\theta_{n,m}$ &the phase shift for device $n$ at the $m^{th}$ reflecting element \\
               \hline
                $\boldsymbol{\Theta}_n$ &the reflection-coefficient matrix for device $n$ \\
               \hline
                $P_n$ &the transmission power of device $n$ \\
               \hline
                $\sigma^2$ &the power of the noise at the receiver \\
               \hline
                $B$ &the uplink bandwidth \\
               \hline
                $\tau_n$ &the transmission latency of device $n$ \\
               \hline
                $\beta_n$ &the binary mode selection variable for device $n$ \\
               \hline

             \end{tabular}
\end{table}
\par In a certain time duration $T$, device $n$ needs to complete a computational task with data size $S_n$. The time duration $T$ shall be less than the channel coherent time. To avoid inter-device interference, we allocate exclusive time intervals for devices for transmission. Binary offloading is considered since it is more practical than partial offloading in the case of inseparability of the application data and easier to implement. Thus, in each time duration, each device either offloads all of its data that need to be processed to the edge server through the wireless link or processes all of its data locally.
\par After the computation is completed, the results are transmitted back from the edge server to the devices. We assume that the edge server has a relatively strong computing capability. Jointly considering the downlink high power of the AP and the very small date sizes of the results, the computing time at the edge server and the downlink transmission time are very small values and negligible. The notations that will be used in the rest of this paper are summarized in Table I.
\subsection{Local Computing Model}
For device $n$, the number of CPU cycles required for processing 1 bit of data is denoted by $C_n$, which depends on the specific computation task. Thus, the total CPU cycles to process the data for device $n$ is $S_nC_n$. Then, we can model the energy consumption of local computing for device $n$ as $\varepsilon_n S_nC_n {f_n}^2$, where $\varepsilon_n$ is a coefficient depending on the chip architecture and $0 \le f_n \le f_{n,max}$ is the CPU computational frequency of device $n$, with $f_{n,max}$ denoting the maximum CPU frequency of device $n$ \cite{4}. In this paper, it is assumed that $T f_{n,max} \ge S_n C_n, \forall n$, which means each device is able to complete the computational task within duration $T$ by using its maximum computing capability.
\subsection{Offloading Model}
The equivalent baseband channels from device $n$ to the server, from device $n$ to the IRS and from the IRS to the server are denoted by $g_n \in \mathbb{C}$, $\textit{\textbf{h}}_n \in \mathbb{C}^{M\times1}$ and $\emph{\textbf{r}} \in \mathbb{C}^{M\times1}$, respectively, where $\mathbb{C}^{a\times b}$ denotes the set of $a \times b$ complex-valued matrices. All the channels are assumed to be flat fading and quasi-static in the time duration $T$ so that they remain constant when devices are offloading their data. In addition, it is assumed that the channel state information (CSI) of all the channels is known.
\par Since the system operates on a time division manner, there is only one device offloading at any time and we can design the phase shifts of the IRS dedicated to each device. Moreover, the phase shifts switching frequency can be up to 5 MHz \cite{14}, making the switching time negligible comparing to $T$. Denote $\boldsymbol{\theta}_n={[\theta_{n,1}, \theta_{n,2},\ldots, \theta_{n,M}]}^T$ as the phase shifts of all the reflecting elements dedicated to device $n$, where $\theta_{n,m} \in [0,2\pi)$ is the phase shift for device $n$ at the $m^{th}$ reflecting element. Besides, we define a diagonal matrix $\boldsymbol{\Theta}_n=$ diag ($\alpha_{1}e^{j\theta_{n,1}}, \alpha_{2}e^{j\theta_{n,2}},\ldots, \alpha_{M}e^{j\theta_{n,M}}$) as the reflection-coefficient matrix for device $n$, where $j$ represents the imaginary unit and $\alpha_{m} \in [0,1]$ denotes the amplitude reflection coefficient of the $m^{th}$ element. In practice, $\alpha_m$ can be measured in advance. Without loss of generality, we assume $\alpha_{m}=1$ for all $m$.
\par With the transmission power of device $n$ denoted by $P_n$ and the transmitted symbol (zero mean and unit variance) of device $n$ denoted by $a_n$, we can express the signal received by the edge server as
\begin{equation}
y_n=({\emph{\textbf{r}}}^{T} \boldsymbol{\Theta}_n \emph{\textbf{h}}_n+g_n) \sqrt{P_n} a_n+\eta,
\end{equation}
where $\eta$ is the additive white Gaussian noise (AWGN) at the receiver with zero mean and variance $\sigma^2$. From (1), the achievable transmission rate of device $n$ is expressed as
\begin{equation}
R_n=B \log_2(1+\frac{{{|\emph{\textbf{r}}}^{T} \boldsymbol{\Theta}_n \emph{\textbf{h}}_n+g_n|}^2 P_n} {\sigma^2}),
\end{equation}
where $B$ is the uplink bandwidth. Therefore, the transmission latency $\tau_n$ is given by $\tau_n=\frac{S_n}{R_n}$ and the energy consumption for offloading the data of device $n$ is $P_n \tau_n$.

\section{Problem Formulation and Transformation}
In this paper, we aim to minimize the devices' total energy consumption by jointly optimizing the binary offloading modes, the CPU frequencies, the offloading powers, the offloading times and the IRS phase shifts for all devices. The energy consumption minimization problem is formulated as:
\begin{gather}
\begin{align}
\mathcal{P}1:\min_{\substack{\{\beta_n\},\{f_n\},\{P_n\},\\ \{\tau_n\},\{\boldsymbol{\theta}_n\}}}& \sum_{n=1}^{N} \Big(\beta_n P_n \tau_n + (1-\beta_n) \varepsilon_n S_nC_n {f_n}^2\Big), \tag{3a}\\
{\rm s.t.}  \qquad & \beta_n \in \{0,1\}, \ \ \,\forall n, \tag{3b}\\
& \sum_{n=1}^{N} \tau_n \le T, \tag{3c}\\
& \frac{S_n C_n}{f_n} \le T, \ \ \, \forall n, \tag{3d}\\
& \tau_n=\frac{\beta_n S_n}{B \log_2(1+\frac{{{|\emph{\textbf{r}}}^{T} \boldsymbol{\Theta}_n \emph{\textbf{h}}_n+g_n|}^2 P_n} {\sigma^2})},  \ \ \,\forall n, \tag{3e}\\
& 0 \le \theta_{n,m} < 2\pi, \ \ \, \forall n,m, \tag{3f}\\
 & 0 \le f_n \le f_{n,max},  \ \ \, \forall n, \tag{3g}\\
 & P_n \ge 0, \ \ \, \forall n \tag{3h}.
\end{align}
\end{gather}
In $\mathcal{P}1$, (3b) is the binary mode selection constraint with $\beta_n=1$ means device $n$ offloads data while $\beta_n=0$ means device $n$ carries out computation locally. (3c) indicates the sum of offloading latencies cannot exceed the time duration $T$. (3d) means the local computation time cannot exceed the time duration $T$. (3e) gives us the formulation of offloading latency and the $\beta_n$ in the numerator is to ensure that a device will have $\tau_n=0$ if it chooses to compute locally. (3f) expresses the range of IRS phase shifts. (3g) and (3h) limit the CPU frequency and transmission power, respectively.
\par Obviously, problem $\mathcal{P}1$ is a nonconvex problem due to constrains (3b) and (3e), thus difficult to solve in general. Furthermore, due to the existence of the binary mode selection variables $\{\beta_n\}$, the objective function (3a) is discontinuous and extremely nonconvex and it usually requires exponential time complexity to find the optimal solution. Thus, $\mathcal{P}1$ is very challenging to solve.
\par From (3e) and (3h), it can be found that for a device who offloads data (i.e., $\beta_n=1$), we can express its offloading power as $P_n=(2^{\frac {S_n} {\tau_n B}}-1)  \frac {\sigma^2} {{{|\emph{\textbf{r}}}^{T} \boldsymbol{\Theta}_n \emph{\textbf{h}}_n+g_n|}^2}$ and $\tau_n > 0$. On the other hand, for a device who computes locally (i.e., $\beta_n=0$), we have $\tau_n = 0$. Thus, the binary mode selection variable can be absorbed into the time allocation variable. Based on the above observation, we can rewrite $\mathcal{P}1$ as the following equivalent problem:
\begin{gather}
\begin{align}
\mathcal{P}2:\qquad \min_{\{\tau_n\},\{\boldsymbol{\theta}_n\},\{f_n\}}  \quad &\sum_{n=1}^{N}  \left(\phi_n(\tau_n) \frac {\sigma^2} {{{|\emph{\textbf{r}}}^{T} \boldsymbol{\Theta}_n \emph{\textbf{h}}_n+g_n|}^2} \qquad \qquad \right. \notag\\&+ (1- \|{\tau_n}\|_0) \varepsilon_n S_nC_n {f_n}^2\Bigg),  \tag{4a}\\
{\rm s.t.}  \qquad & \tau_n \ge 0,\quad \forall n,  \tag{4b}\\
 & \sum_{n=1}^{N} \tau_n \le T, \tag{4c}\\
 & \frac{S_n C_n}{f_n} \le T, \quad \forall n, \tag{4d}\\
 & 0 \le \theta_{n,m} < 2\pi, \quad \forall n,m, \tag{4e}\\
 & 0 \le f_n \le f_{n,max},  \quad \forall n. \tag{4f}
\end{align}
\end{gather}
where $$ \phi_n(\tau_n)=\left\{
\begin{aligned}
&0, \qquad \qquad \qquad  \ \; \, {\rm when} \ \tau_n=0, \\
&\tau_n (2^{\frac {S_n} {\tau_n B}}-1)  ,\quad \ \,\,{\rm when} \ \tau_n \neq 0,
\end{aligned}
\right.
$$
and $\|\cdot\|_0$ denotes the $\ell_0$-norm.
\par From $\mathcal{P}2$, it can be seen that for a device with $\tau_n=0$, its energy consumption depends on its CPU frequency. On the other hand, for a device with $\tau_n>0$, its energy consumption depends on $\tau_n$ and the dedicated reflecting elements vector. Thus, we need to allocate $\tau_n$ (could be zero, if computing locally) to each device, optimize the dedicated reflecting elements vector for each device whose $\tau_n$ does not equal to zero and optimize the CPU frequency for each device whose $\tau_n$ equals to zero.
\par However, it can be observed from (4a) that $\{\boldsymbol{\theta}_n\}$ and $\{f_n\}$ are fully separable with respect to $n$, therefore, they can be solved individually. In particular, each optimal $f_n$ is $\frac {S_n C_n} {T}$, which is the minimum CPU frequency satisfying constraint (4d). Furthermore, each optimal $\boldsymbol{\theta}_n$ is the $\boldsymbol{\theta}_n$ which maximizes ${{{|\emph{\textbf{r}}}^{T} \boldsymbol{\Theta}_n \emph{\textbf{h}}_n+g_n|}^2}$. Denoting the $m^{th}$ elements of $\emph{\textbf{r}}$ and $\emph{\textbf{h}}_n$ as $r_m$ and $h_{n,m}$, respectively, we have $|{\emph{\textbf{r}}}^{T} \boldsymbol{\Theta}_n \emph{\textbf{h}}_n+g_n|=|\sum_{m=1}^M r_m e^{j\theta_{n,m}} h_{n,m}+g_n|$. To maximize this norm, we need $\hat\theta_{n,m}=mod\Big(2\pi,phase(g_n)-phase(r_m  h_{n,m})\Big)$. From this, it is clear to see the function of the IRS: to control the environment smartly, making every equivalent reflecting channel combines coherently with the direct channel $g_n$, thus maximizing the channel gain of the equivalent composite channel. This insight is brought by the availability of dedicated design to phase shifts of the IRS for each device.
\par After determining $\{\boldsymbol{\theta}_n\}$ and $\{f_n\}$, $\mathcal{P}2$ becomes:
\begin{gather}
\begin{align}
\mathcal{P}3: \qquad \min_{\{\tau_n\}} \quad & \sum_{n=1}^{N} \left(b_n \phi_n(\tau_n) + (1- \|{\tau_n}\|_0) \frac{\varepsilon_n S_n^3 C_n^3} {T^2}\right),  \tag{5a}\\
{\rm s.t.}  \qquad &\tau_n \ge 0,\quad \forall n, \tag{5b}\\
&\sum_{n=1}^{N} \tau_n \le T,\tag{5c}
\end{align}
\end{gather}
where $b_n=\frac {\sigma^2}{{{|\emph{\textbf{r}}}^{T} {\hat{\boldsymbol{\Theta}}_n} \emph{\textbf{h}}_n+g_n|}^2}$ and ${\hat{\boldsymbol{\Theta}}_n}$ denotes the optimal $\boldsymbol{\Theta}_n$. Although there is only one set of variables $\{\tau_n\}$ and the feasible set becomes a convex set, this problem is still challenging due to the multiple discontinuities of the objective function brought by the piecewise functions $\{\phi_n(\tau_n)\}$ and the $\ell_0$-norm.
\section{Greedy-Based Algorithm}
To tackle the discontinuities in the objective function, we first assume the knowledge of which device has $\tau_n=0$ and which device has $\tau_n>0$. With the set of devices with $\tau_n=0$ denoted by $\mathcal{U}_l$ and the set of devices with $\tau_n>0$ as $\mathcal{U}_o$, $\mathcal{P}3$ becomes
\begin{gather}
\begin{align}
\mathcal{P}4: \quad \ \,\,\min_{\{\tau_n\}} \quad & \sum_{n \in \mathcal{U}_o}  b_n \tau_n (2^{\frac {S_n} {\tau_n B}}-1) + \sum_{n \in \mathcal{U}_l} \frac{\varepsilon_n S_n^3 C_n^3} {T^2}, \tag{6a}\\
{\rm s.t.}  \qquad & \tau_n > 0,\quad \forall n \in \mathcal{U}_o, \tag{6b}\\
  & \sum_{n \in \mathcal{U}_o} \tau_n \le T. \quad  \tag{6c}
\end{align}
\end{gather}
Next, a fundamental characteristic of $\mathcal{P}4$ is given in the following $Lemma \ 1$, which is proved in Appendix A.
\par $Lemma \ 1:$ Problem $\mathcal{P}4$ is a convex optimization problem and the function $b_n \tau_n (2^{\frac {S_n} {\tau_n B}}-1)$ is a monotonically decreasing function with respect to $\tau_n>0$.
\par Since $\mathcal{P}4$ is a convex optimization problem, general convex optimization techniques, such as interior point method, can be employed to solve it. However, such method has a high computational complexity. Therefore, we propose a low-complexity bisection search based algorithm as follows.
\par Firstly, we can assert that the optimal solution of $\mathcal{P}4$ must make inequality constraint (6c) take the equality, since the objective function is a monotonic decreasing function with respect to any $\tau_n$ in the feasible set according to $Lemma \ 1$. By introducing a Lagrangian multiplier to constraint $\sum\limits_{n \in \mathcal{U}_o} \tau_n=T$, a partial Lagrangian function is constructed:
\begin{equation}
L(\{\tau_n\}_{n \in \mathcal{U}_o},\nu)=\sum_{n \in \mathcal{U}_o} b_n \tau_n (2^{\frac {S_n} {\tau_n B}}-1)+\nu (\sum_{n \in \mathcal{U}_o} \tau_n-T). \tag{7}
\end{equation}
Then, we can apply KKT conditions to obtain the sufficient and necessary conditions of the optimal solution:
\begin{gather}
\begin{align}
\frac {\partial L} {\partial {\tau_n}^*}=b_n \Big(2^{\frac {S_n} {{\tau_n}^* B}} & (1-\ln2 \frac {S_n} {{\tau_n}^* B} )-1\Big)+ \nu^*=0, \quad \forall n \in \mathcal{U}_o \tag{8a}\\
 & \sum_{n \in \mathcal{U}_o} \tau_n^* = T. \tag{8b}
\end{align}
\end{gather}
Note that the optimal solution still need to satisfy $\tau_n^* > 0, \quad \forall n \in \mathcal{U}_o$. Define $\psi_n(\tau_n)=b_n\Big(2^{\frac {S_n} {{\tau_n} B}}  (1-\ln2 \frac {S_n} {{\tau_n} B})-1\Big)$ with domain $\tau_n > 0$, then, according to $Lemma \ 1$, we know that $\psi_n(\tau_n)$ is a  monotonically increasing function and its value must be negative.
\par Based on this knowledge, with a given positive $\nu$, we have that $\sum\limits_{n \in \mathcal{U}_o} \psi_n^{-1}(-\nu)$ is a monotonically decreasing function with respect to $\nu$. If this value is smaller than $T$, we can assert that the optimal $\nu^*$ is smaller than this $\nu$. On the contrary, if this value is larger than $T$, it can be concluded that the optimal $\nu^*$ is larger than this $\nu$. Thus, a bisection search algorithm can be used to find the $\nu^*$ in (8a). After obtaining the $\nu^*$, we have $\tau_n^*=\psi_n^{-1}(-\nu^*), \forall n \in \mathcal{U}_o$. Comparing to the conventional interior point method with the computational complexity of $\mathcal{O}(|\mathcal{U}_o|^3)$ where $|\mathcal{U}_o|$ denotes the number of elements in $\mathcal{U}_o$ \cite{20}, the bisection search algorithm has a $\mathcal{O}(|\mathcal{U}_o|)$ complexity.
\par With $\mathcal{P}4$ readily solved, the remaining challenge is to determine whether each device should take $\tau_n=0$ or $\tau_n>0$. While exhaustively enumerating all possibilities generates the optimal solution, such method would cost a high complexity with $\mathcal{O}(2^N)$. Therefore, we propose a greedy-based method to deal with this problem. That is, we start at $\mathcal{U}_o=\{\}$, which means all devices compute locally. Then, we choose the "best" device to offload data among the devices who have not been chosen to offload, i.e., changing the corresponding $\tau_n=0$ to $\tau_n>0$. The "best"  means choosing the device (among the devices currently not offloading their data) that leads to the maximum decrease in the system energy consumption (i.e., objective function (6a)). If the energy consumption cannot be reduced in a particular iteration, we end the greedy-based algorithm and the current $\{\tau_n\}$ is the final result. The whole greedy-based algorithm is summarized in Algorithm 1.
\par It is evident that the computational complexity to execute step 5 of Algorithm 1 for a particular $k\in\mathcal{U}_l$ is linear with respect to the number of devices, namely, $\mathcal{O}(N)$. Under the worst situation, we need to repeat such operation for $\sum\limits_{n=1}^N n=\frac {N^2+N}{2}$ times so the overall computational complexity is $\mathcal{O}(N^3)$, which is significantly lower than that of the enumeration method. However, this is still unsatisfactory in large-scale devices systems. In order to further reduce the computational complexity, a penalty-based algorithm is proposed in the next section.
\begin{algorithm}
\caption{Greedy-Based Algorithm for $\mathcal{P}2$}
\label{}
\begin{algorithmic}[1]
\STATE \textbf{initialize}  $\mathcal{U}_o=\{\}$, $\mathcal{U}_l=\{1,2,...,N\}$, $E_o=0$, $E_l =\sum\limits_{n=1}^{N} \frac{\varepsilon_n S_n^3 C_n^3} {T^2}$.\\
\STATE $\hat\theta_{n,m} = mod\Big(2\pi,phase(g_n)-phase(r_m  h_{n,m})\Big), \ \forall m,n$.\\
\STATE $b_n =\frac {\sigma^2}{{{|\emph{\textbf{r}}}^{T} {\hat{\boldsymbol{\Theta}}_n} \emph{\textbf{h}}_n+g_n|}^2}, f_n=\frac{S_nC_n}{T}, \ \forall n$.\\
\STATE \textbf{repeat}
\STATE \quad find $k\in\mathcal{U}_l$ such that $ED=E_o+E_l-\left(\sum\limits_{n \in \mathcal{U}_o \cup k} b_n  \tau_n^* \right.$ \\ \quad $\left. (2^{\frac {S_n} {\tau_n^* B}}-1)+\sum\limits_{n \in \mathcal{U}_l \setminus k} \frac{\varepsilon_n S_n^3 C_n^3} {T^2}\right)$ is maximized, where \\ \quad $\tau_n^*=\psi_n^{-1}(-\nu^*)$ and $\nu^*$ is found by applying bisection \\ \quad search method on $\sum\limits_{n \in \mathcal{U}_o \cup k} \psi_n^{-1}(-\nu)=T$.\\
\STATE \quad  \textbf{if} $ED>0 \ \textbf{then}$\\
\STATE \quad \ \ $E_o=\sum\limits_{n \in \mathcal{U}_o \cup k} b_n \tau_n^* (2^{\frac {S_n} {\tau_n^* B}}-1)$, $E_l=\sum\limits_{n \in \mathcal{U}_l \setminus k} \frac{\varepsilon_n S_n^3 C_n^3} {T^2}$, \\ \quad \ \ \,$\mathcal{U}_o=\mathcal{U}_o \cup k$ and $\mathcal{U}_l=\mathcal{U}_l \setminus k$.\\
\STATE \quad  \textbf{else}\\
\STATE \quad \ \ \textbf{break}\\
\STATE \textbf{return} $\{\tau_n\},\{\boldsymbol{\theta}_n\},\{f_n\}$.
\end{algorithmic}
\end{algorithm}
\section{Penalty-Based Algorithm}
\par The advantages of the greedy-based algorithm is that it is intuitive and much simpler than enumeration method. However, it cannot fit the large-scale devices context well. Different from the greedy-based algorithm, the penalty-based algorithm determines offloading or not for all devices in parallel, thus significantly reduces the computational complexity to $\mathcal{O}(N)$.
\par Although $\mathcal{P}3$ is discontinuous, it is noticed that the objective function is naturally decoupled with respect to $\{\tau_n\}$ and they are only coupled in constraint (5c). To decompose this problem into $N$ parallel subproblems, we first introduce $N$ artificial variables $\{a_n\}$ and equivalently reformulate $\mathcal{P}3$ as:
\begin{gather}
\begin{align}
\mathcal{P}5: \qquad \min_{\{\tau_n\},\{a_n\}} \ & \sum_{n=1}^{N} \left(b_n\phi_n(a_n) + (1- \|{a_n}\|_0) \frac{\varepsilon_n S_n^3 C_n^3} {T^2}\right) \notag\\ &+ l\left(\{\tau_n\}\right), \tag{9a}\\
 {\rm s.t.}  \qquad & a_n \ge 0,\quad \forall n, \tag{9b}\\
 & \tau_n=a_n, \quad \forall n,\tag{9c}
\end{align}
\end{gather}
where functions $\{\phi_n\}$ are defined in Section III and
$$ \qquad \quad \,l\left(\{\tau_n\}\right)=\left\{
\begin{aligned}
&0,\ \ {\rm if} \ \sum_{n=1}^{N} \tau_n \le T \ \ {\rm and} \ \ \tau_n \ge 0, \ \forall n, \\
& \infty, \ {\rm otherwise}.
\end{aligned}
\right.
$$
One might be tempted to introduce Lagrangian multipliers associated to constraint (9c) to construct a partial Lagrangian function, and then apply augmented Lagrangian or Alternating Direction Method of Multipliers (ADMM). However, due to the discontinuities of the objective function, the convergence of ADMM cannot be guaranteed. Besides, even though the dual problem in augmented Lagrangian is guaranteed to converge, due to the non-convexity of this problem, there is a dual-gap which cannot be estimated precisely. Therefore, instead of introducing Lagrangian multipliers, we propose a penalty-based algorithm to solve $\mathcal{P}5$.
\par By introducing a nonnegative penalty factor $\rho$ associated with constraint (9c) and putting it into the objective function, $\mathcal{P}5$ becomes:
\begin{gather}
\begin{align}
\mathcal{P}6: \qquad \min_{\{\tau_n\},\{a_n\}}  \ & \sum_{n=1}^{N} \left(b_n\phi_n(a_n) + (1- \|{a_n}\|_0) \frac{\varepsilon_n S_n^3 C_n^3} {T^2}\right)\notag\\ &+ l\left(\{\tau_n\}\right) +\rho \sum_{n=1}^{N} (\tau_n-a_n)^2, \tag{10a}\\
 {\rm s.t.}  \qquad & a_n \ge 0,\quad \forall n, \tag{10b}
\end{align}
\end{gather}
where $\rho$ is to control the degree of matching between the original problem $(P5)$ and the penalized problem $(P6)$. When $\rho \rightarrow \infty$, $(P6)$ is exactly the same as $(P5)$ since we can regard the term $\rho \sum_{n=1}^{N} (\tau_n-a_n)^2$ as an indicator function associated to the constraint (9c). On the other hand, when $\rho \rightarrow 0$, the constraint (9c) will be relaxed.
\par Since $\mathcal{P}6$ depends on two block of variables, the block coordinate descent (BCD) technique, which iteratively solves $\{a_n\}$ with $\{\tau_n\}$ fixed and solves $\{\tau_n\}$ with $\{a_n\}$ fixed, is employed. In particular, with $\{\tau_n\}$ fixed, the subproblem for each of the $a_n$ is
\begin{gather}
\begin{align}
\mathcal{P}7:\ \min_{a_n} \quad  &  b_n \phi_n(a_n) + (1- \|{a_n}\|_0) \frac{\varepsilon_n S_n^3 C_n^3} {T^2}+\rho (\tau_n-a_n)^2, \tag{11a}\\
{\rm s.t.}  \qquad & a_n \ge 0. \tag{11b}
\end{align}
\end{gather}
\par Benefitting from the decomposition to $N$ parallel smaller problems, the objective function of $\mathcal{P}7$ is only discontinuous at a single point $a_n=0$ so that this problem can be solved by substituting $a_n=0$ or $a_n>0$, and compare their respective objective values. For $a_n=0$, the value of the objective function is $\frac{\varepsilon_n S_n^3 C_n^3} {T^2}+\rho \tau_n^2$. On the other hand, for $a_n>0$, we need to minimize $b_n a_n (2^{\frac {S_n} {a_n B}}-1)+\rho (\tau_n-a_n)^2$. According to Appendix A, $b_n a_n (2^{\frac {S_n} {a_n B}}-1)$ is a convex function with respect to $a_n>0$. Also, it is obvious that $\rho (\tau_n-a_n)^2$ is a convex function since it is a concave parabola. Therefore, minimizing $b_n a_n (2^{\frac {S_n} {a_n B}}-1)+\rho (\tau_n-a_n)^2$ with respect to $a_n>0$ is a one-dimension convex optimization problem so that the optimal solution can be obtained by using bisection search method to find the zero point of its derivative function $b_n\Big(2^{\frac {S_n} {{a_n} B}}  (1-\ln2 \frac {S_n} {{a_n} B})-1\Big)- 2 \rho (\tau_n-a_n)$.
\begin{algorithm}
\caption{Projected Gradient Descent for $\mathcal{P}8$}
\label{}
\begin{algorithmic}[1]
\STATE \textbf{initialize} ${\boldsymbol{\tau}=[\tau_1, \tau_2,\ldots, \tau_N]}^T$ such that $\tau_n \ge 0, \ \forall n$ and $\sum_{n=1}^{N} \tau_n \le T$\\
\STATE \textbf{repeat} \\
\STATE  \quad update $\boldsymbol{t}=\boldsymbol{\tau}-\kappa \nabla k(\boldsymbol{\tau})$, where $k(\boldsymbol{\tau})=\sum_{n=1}^{N} ({\tau_n}^2-$ \\ \quad $2 a_n \tau_n)$ and $\kappa$ is the step size\\
\STATE \quad \textbf{if} $t_n \ge 0, \ \forall n$ and $\sum_{n=1}^{N} t_n \le T$ \textbf{then}\\
\STATE \quad \ \;$\boldsymbol{\tau}=\boldsymbol{t}$\\
\STATE \quad \textbf{else}\\
\STATE \quad \ \ use bisection search method to find $\mu$ satisfying \\ \quad \ \ $\sum_{n=1}^{N} {(t_n-\mu)}_+=T$, where $(x)_+=max\{x,0\}$ \\ \quad \ \ and $\boldsymbol{\tau}={(\boldsymbol{t}-\mu)}_+$\\
\STATE \textbf{until} stopping criterion is satisfied
\STATE \textbf{return} $\{\tau_n\}$
\end{algorithmic}
\end{algorithm}
\begin{algorithm}
\caption{Penalty-Based Algorithm for $\mathcal{P}2$}
\label{}
\begin{algorithmic}[1]
\STATE \textbf{initialize} $\tau_n = \frac {T}{N}, \ \forall n$ and $\rho=300$\\
\STATE $\hat\theta_{n,m} = mod\Big(2\pi,phase(g_n)-phase(r_m  h_{n,m})\Big), \ \forall m,n$. \\
\STATE $b_n =\frac {\sigma^2}{{{|\emph{\textbf{r}}}^{T} {\hat{\boldsymbol{\Theta}}_n} \emph{\textbf{h}}_n+g_n|}^2}, f_n=\frac{S_nC_n}{T}, \ \forall n$.\\
\STATE \textbf{repeat} \\
\STATE \quad \textbf{for} $n=1,2,...,N$\\
\STATE \quad \ \, \textbf{if} $b_n a_n^* (2^{\frac {S_n} {a_n^* B}}-1)+\rho (\tau_n-a_n^*)^2<\frac{\varepsilon_n S_n^3 C_n^3} {T^2}+\rho \tau_n^2$, \\ \quad \,\,\,\,\,where $a_n^*$ is obtained by using bisection search \\ \quad \,\,\,\,\,method to find the solution of $b_n\Big(2^{\frac {S_n} {{x} B}}  (1-\ln2 \frac {S_n} {{x} B})$\\$\quad \,\,\,\,\, -1\Big)- 2 \rho (\tau_n-x)=0$ with respect to $x>0$, \textbf{then}\\
\STATE \quad \quad \,\,\,\,$a_n=a_n^*$\\
\STATE \quad \ \, \textbf{else}\\
\STATE \quad \quad \,\,\,\,$a_n=0$\\
\STATE \quad \textbf{end for}\\
\STATE\quad update $\{\tau_n\}$ by using Algorithm 2\\
\STATE \textbf{until} stopping criterion is satisfied\\
\STATE \textbf{return} $\{\tau_n\},\{\boldsymbol{\theta}_n\},\{f_n\}$
\end{algorithmic}
\end{algorithm}
\par On the other hand, when $\{a_n\}$ are fixed, $\mathcal{P}6$ becomes
\begin{gather}
\begin{align}
\mathcal{P}8: \qquad  \,\,\, \ \ \quad \ \ \ \min_{\{\tau_n\}} \quad  & \sum_{n=1}^{N} ({\tau_n}^2-2 a_n \tau_n), \qquad \qquad \tag{12a} \\
{\rm s.t.}  \qquad & \tau_n \ge 0, \quad \forall n,\tag{12b} \\
&\sum_{n=1}^{N} \tau_n \le T. \tag{12c}
\end{align}
\end{gather}
Fortunately, $\mathcal{P}8$ is also a convex optimization problem since the objective function is just the sum of concave parabolas and the constraint functions are affine functions. Furthermore, since the feasible set is a simplex, we can use the projected gradient descent (PGD) method, which is summarized in Algorithm 2, instead of the conventional interior point method to solve this problem in order to reduce the computational complexity. Since $\mathcal{P}7$ and $\mathcal{P}8$ can be optimally solved, BCD guarantees a critical point upon convergence. Finally, the whole penalty-based algorithm is summarized in Algorithm 3.
\par For the complexity of $\mathcal{P}7$ and $\mathcal{P}8$, it can be seen that $\mathcal{P}7$ is a series of $N$ parallel one-dimensional convex optimization problems, so the complexity for solving $\mathcal{P}7$ is linear with respect to $N$. Furthermore, $\mathcal{P}8$ is a $N$-dimensional convex optimization problem and is solve by the PGD method which is a first-order method with $\mathcal{O}(N)$ complexity, thus, the complexity for solving $\mathcal{P}8$ is also linear with respect to $N$. Consequently, the whole penalty-based algorithm has a computational complexity of $\mathcal{O}(N)$, which is significantly lower than that of the greedy-based algorithm.
\section{Extension to Discrete Phase Shifts}
In Section II, we adopt the model of continuous IRS phase shifts, where each reflecting element is capable of taking an arbitrary phase shift value in the interval of $[0,2\pi)$. However, in practice, only a limited number of discrete IRS phase shift values may be available \cite{30}. Specifically, the phase shift values at each reflecting element is constrained to the set $\mathcal{F}=\{0, 2\pi/L,..., (L-1)2\pi/L\}$, where $L$ denotes the number of total phase shift values.
\par Under this context, the phase shift design problem becomes $\max\limits_{\theta_{n,m} \in \mathcal{F}} |\sum_{m=1}^M r_m e^{j\theta_{n,m}} h_{n,m}+g_n|$. Unfortunately, since different $\theta_{n,m}$ are coupled inside the absolute value function, the optimal $\{\theta_{n,m}\}_{n=1}^N$ can only be obtained by exhaustive search, which has a high complexity of $\mathcal{O}(L^M)$, and is prohibitive for large $M$. Thus, we propose to first quantize the optimal continuous phase shift result to its nearest discrete phase value in $\mathcal{F}$. Mathematically, we have
\begin{equation}
\widetilde\theta_{n,m}=\arg \min_{\theta \in \mathcal{F}} \left|\theta-mod\Big(2\pi,phase(g_n)-phase(r_m  h_{n,m})\Big)\right|. \tag{13}
\end{equation}
Then, with a given $n$, an alternating maximization method is employed to optimize each $\theta_{n,m}$ with $m=1,2,\ldots,M$. Specifically, with the obtained ${[\widetilde\theta_{n,1}, \widetilde\theta_{n,2},\ldots, \widetilde\theta_{n,M}]}^T$ as the initialized value, we alternatively update $\max\limits_{\theta_{n,k} \in \mathcal{F}} |\sum_{m=1}^M r_m e^{j\theta_{n,m}} h_{n,m}+g_n|$ with respect to $\theta_{n,k}$ while holding $\{\theta_{n,m}\}$ with $m \neq k$ fixed. Since the objective function value is non-decreasing in each iteration, repeatedly executing this from $k=1$ to $k=M$ until convergence guarantees the final solution is better than ${[\widetilde\theta_{n,1}, \widetilde\theta_{n,2},\ldots, \widetilde\theta_{n,M}]}^T$ which is obtained by simple quantization. Finally, the convergence of this alternating maximization method is guaranteed since the optimal value of the objective function is upper-bounded.
\par Thus, for the scenario of discrete IRS phase shifts, we replace the second steps in Algorithm 1 and Algorithm 3 with the above procedure. Due to the constraint of discrete IRS phase shifts, a performance loss is expected compared to the continuous phase model. In simulations, the impact will be investigated.
\section{Simulation Results and Discussions}
In this section, numerical results are provided to demonstrate the benefits of the IRS-aided binary offloading MEC system and the proposed algorithms. We consider a multi-device system with eight devices denoted by $D_1, D_2,\ldots, D_8$, respectively. As illustrated in Fig. 2, the edge server (the antenna of the AP), the eight devices and the IRS are located on different heights with respect to the base plane, which are 10 m, 0 m and 5 m, respectively. Furthermore, $D_1$, $D_2$, $D_3$ and $D_4$ lie randomly on a circle centered at the projection of the edge server with a radius $d_1=20$ m and the other four devices $D_5$, $D_6$, $D_7$ and $D_8$ lie randomly on a half circle centered at the projection of the IRS with a radius $d_2=3$ m. In the simulations, we consider the homogeneous computing configuration, which means all the $\varepsilon_n$, $S_n$, $C_n$ and $f_{n,max}$ are the same for different $n$ and they are set as $\varepsilon_n=10^{-28}$, $S_n=1$ MB, $C_n=100$ cycle/bit and $f_{n,max}=1$ GHz.
\begin{figure}[htbp]
\centering
\includegraphics[width=3.5in]{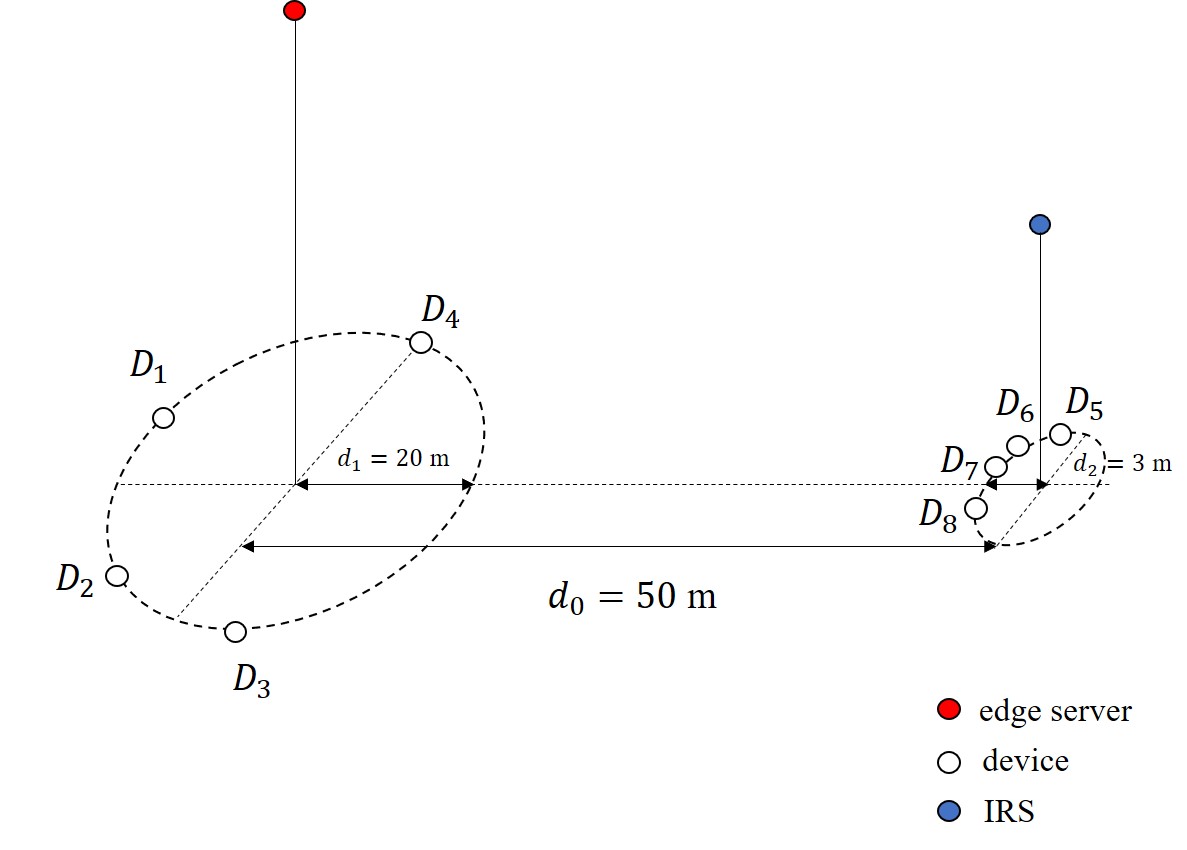}
\caption{The system setup in the simulations}
\end{figure}
\par For signal attenuation, the path loss model is given by $L(d)=\lambda \left(\frac {d} {D_0}\right)^{-\alpha}$, where $\lambda$ is the path loss at the reference distance $D_0=1$ m, $d$ denotes the link distance and $\alpha$ denotes the path loss exponent. In the simulations, we set $\lambda=10^{-3}$, $\alpha=3$, and it is assumed that the device-server channel and the device-IRS channels are Rayleigh fading. In practice, the IRS is usually deployed with the knowledge of the AP's location to exploit LoS channel so the IRS-server channel is assumed to be Rician fading with Rician factor $K=20$ dB. Other parameters in the system are set as follows: $B=10$ MHz, $\sigma^2=10^{-10}$ W and $T=1$ s. All programs are run on MATLAB R2020b by a Windows X64 laptop equipped with Intel Core i7-7700 HQ CPU and 8 GB RAM and each point in simulation is averaged over 300 simulation trials with independent realizations of location of devices, channels and noise.
\subsubsection{Performance versus Number of Reflecting Elements}
In Fig. 3, we illustrate the total energy consumption of devices versus the number of reflecting elements under different schemes. In order to clearly see the effectiveness of the proposed algorithms, we use the enumeration method which can be guaranteed to get the optimal solution by enumerating all $2^8$ possibilities as a comparison. It can be discovered that both the greedy-based and the penalty-based achieve the near-optimal solutions. For the effect of IRS, it can be seen clearly that no matter which algorithm is used, employing IRS reduces the total energy consumption compared to the case without IRS, and the saving of total energy due to IRS is more and more prominent when the number of reflecting elements increases. Furthermore, the greedy-based algorithm achieves slightly better performance than that of the penalty-based algorithm. For the greedy-based algorithm, it determines which device should offload one by one. More specifically, under a particular offloading decision, it looks for a device that leads to a maximum reduction of energy consumption by switching it from computing locally to offloading. Thus, it considers the overall situation of energy consumption more holistically. On the other hand, in order to reduce the computational complexity, the penalty-based algorithm decomposes the original problem into many parallel subproblems so that the offloading decision for each device is determined in parallel. \begin{figure}[htbp]
\centering
\includegraphics[width=3.5in]{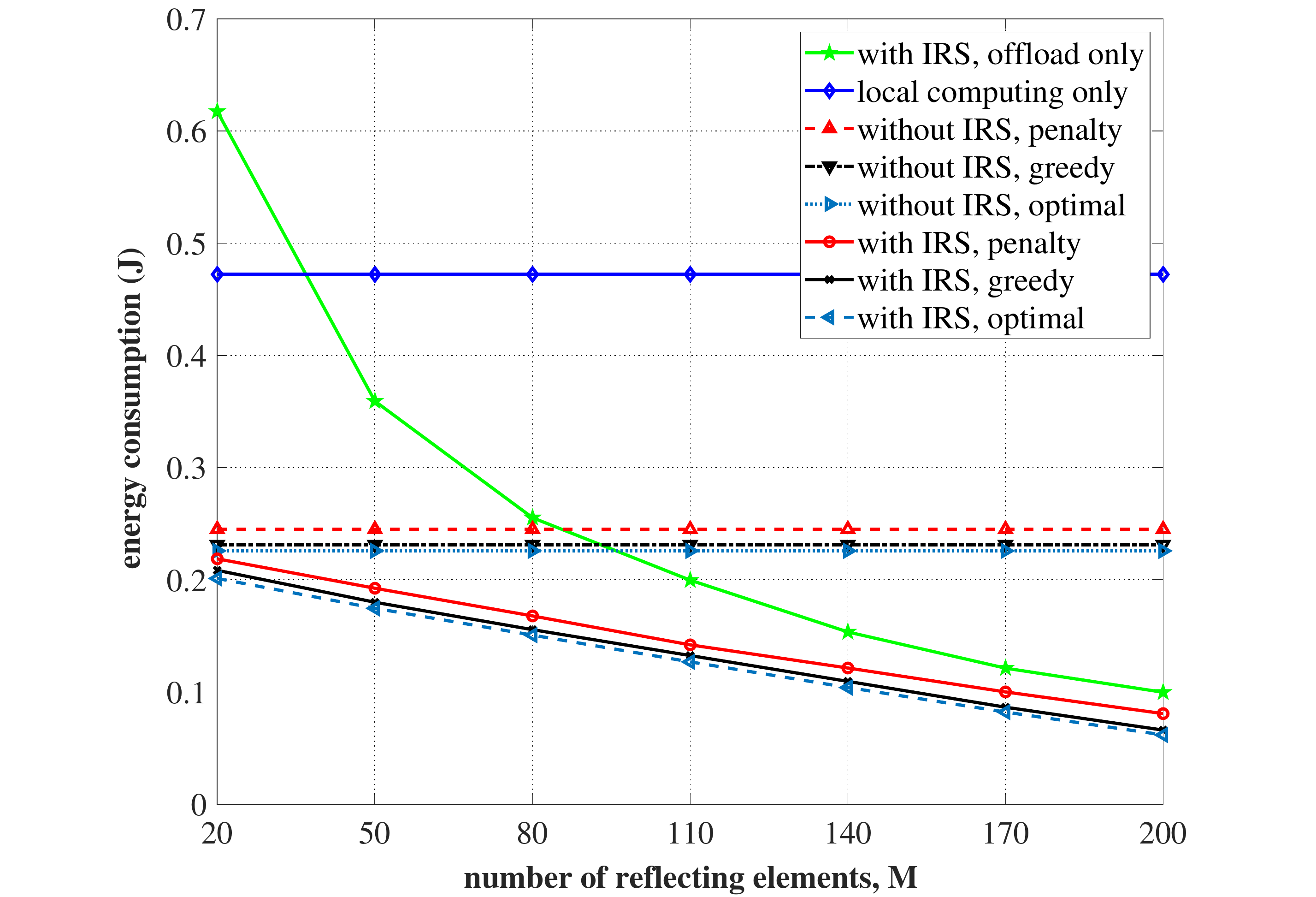}
\caption{Energy Consumption versus Number of Reflecting Elements}
\end{figure}
\begin{figure}[htbp]
\centering
\includegraphics[width=3.5in]{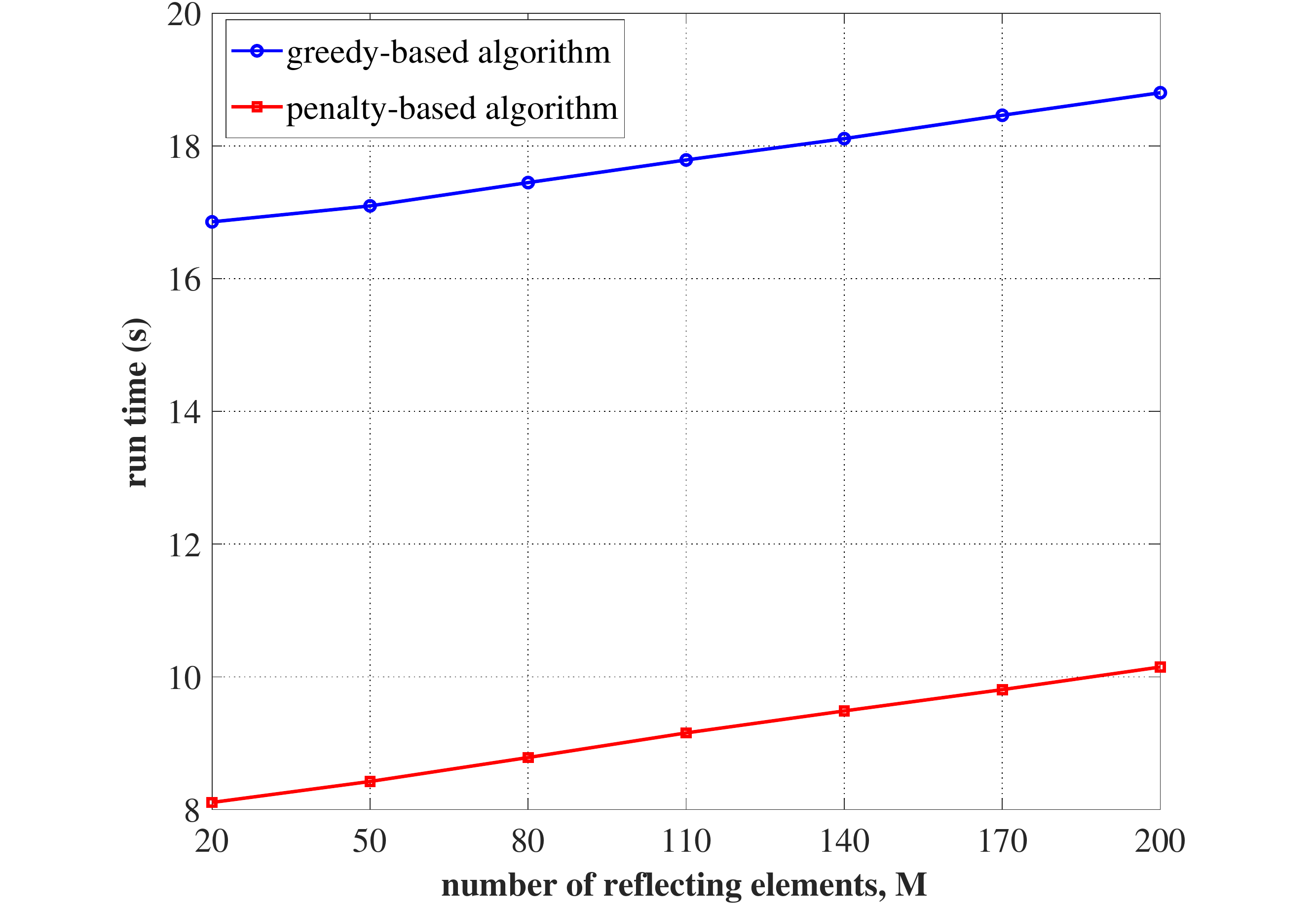}
\caption{Run Time versus Number of Reflecting Elements}
\end{figure}Since the decision at each device is made in parallel, it is reasonable to see it performs not as well as greedy-based algorithm. In practice, trade-off between performance and computational complexity shall be considered.
\par Besides, by employing a large number of reflecting elements ($M\geq 110$), even the case of all devices offloading achieves a better performance than that of executing binary offloading but without IRS. Furthermore, with the increase of the number of reflecting elements, the difference between employing IRS with all devices offloading and employing IRS with binary offloading (both greedy and penalty) becomes smaller and smaller. The reason is that with the increase of the number of reflecting elements, devices have larger probabilities to offload due to the enhancement of the capability of the IRS. In fact, when the number of the reflecting elements is large enough, making all devices offload achieves the near optimal solution since each device will offload with probability close to 1 due to the dramatically enhanced channel quality.
\begin{figure}[htbp]
\centering
\includegraphics[width=3.5in]{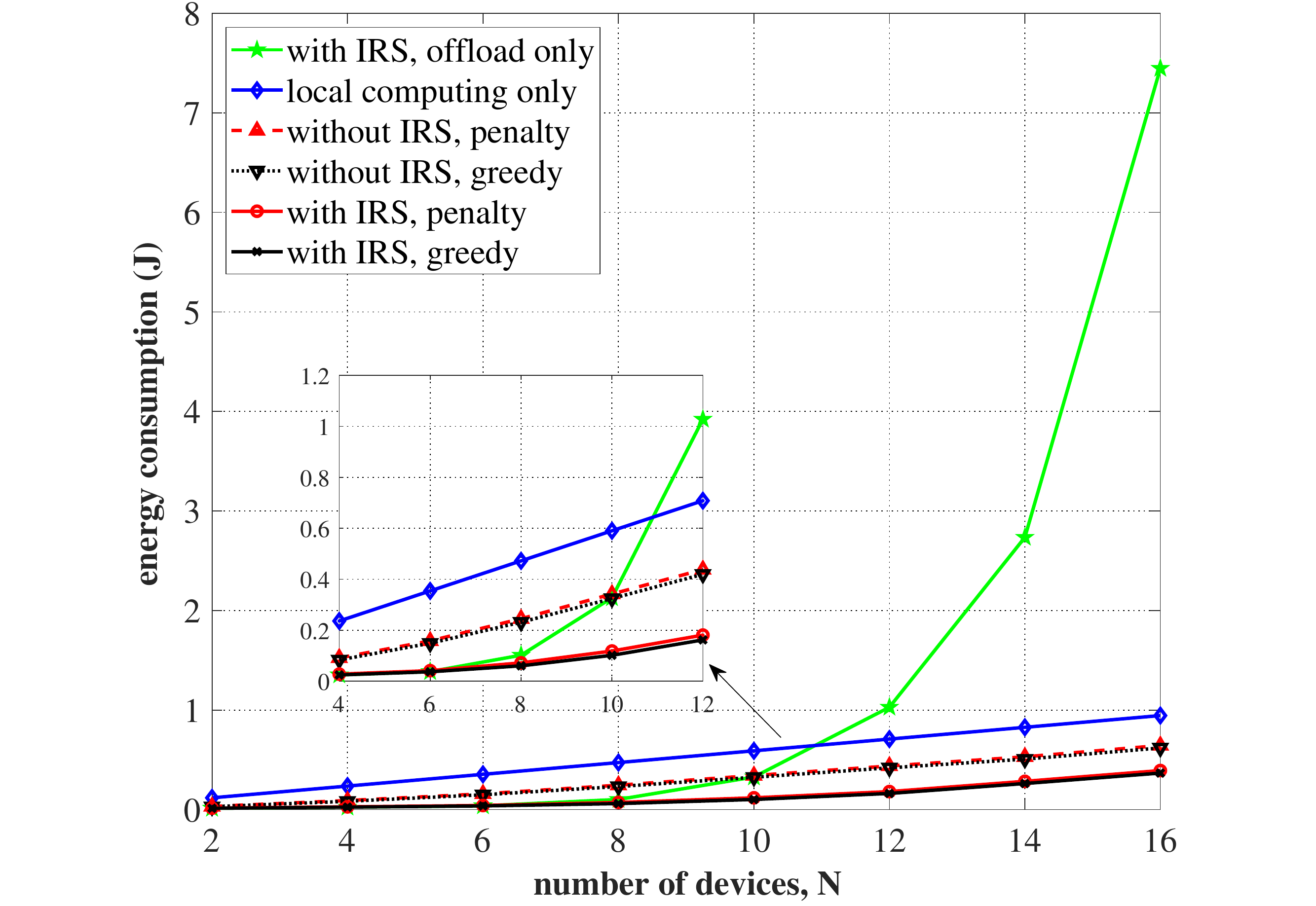}
\caption{Energy Consumption versus Number of Devices}
\end{figure}
\begin{figure}[htbp]
\centering
\includegraphics[width=3.5in]{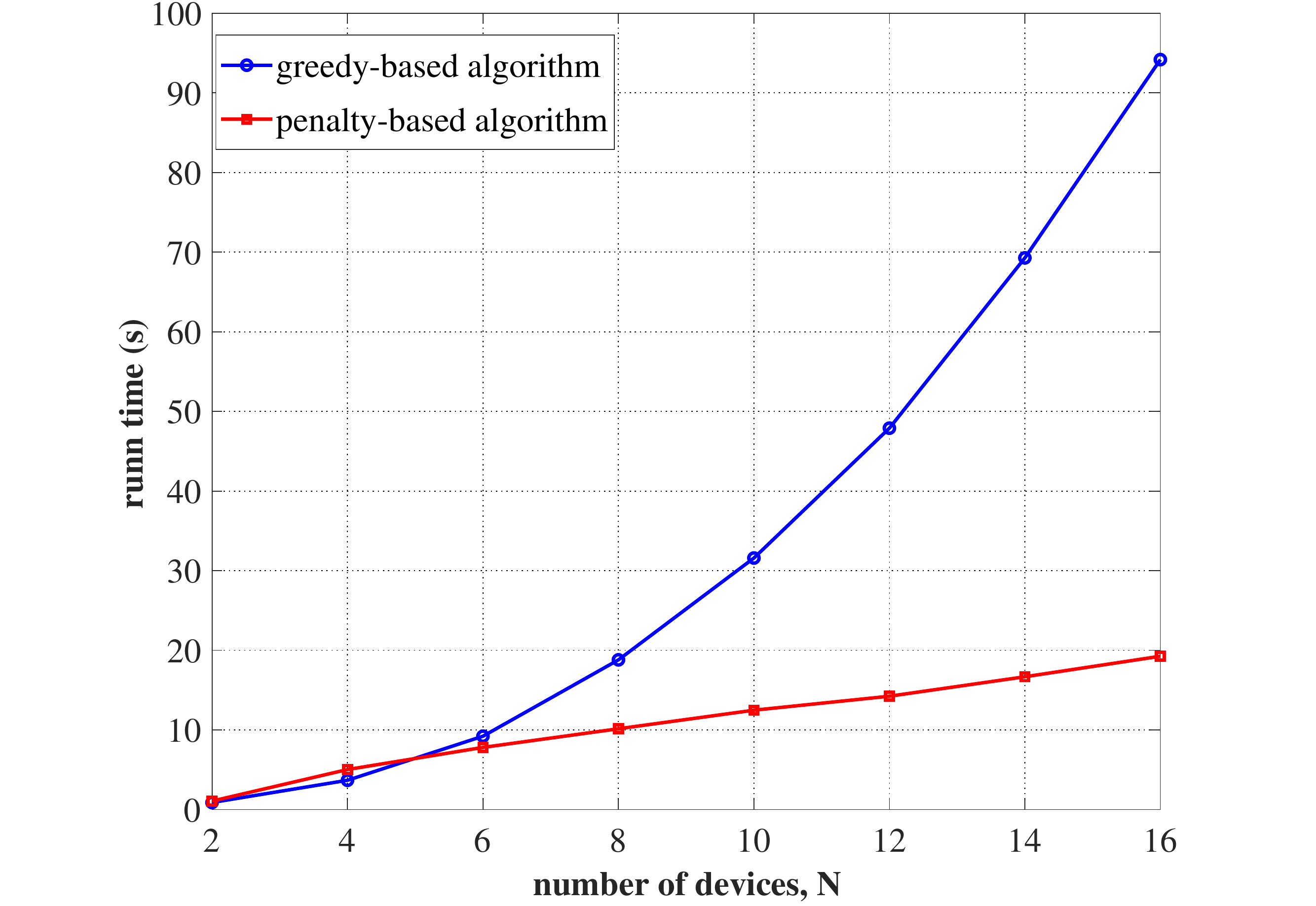}
\caption{Run Time versus Number of Devices}
\end{figure}
\par Fig. 4 shows the run times of greedy and penalty-based algorithms versus the number of reflecting elements. The system setup is the same as that for generating Fig. 3.  It can be seen that the penalty-based algorithm saves near half of the computation time compared to the greedy-based algorithm. Furthermore, with the increase of the number of reflecting elements, the rate of increases in run times of both of the algorithms are linear with almost identical slopes. This is because the two algorithms solve the optimal IRS phase shifts via the same procedure (step 2 in Algorithm 1 and step 2 in Algorithm 3) and the computational complexity of this procedure is linear with respect to the number of reflecting elements.
\subsubsection{Performance versus Number of Devices}
Fig. 5 illustrates the total energy consumption of devices versus the number of devices. The system setup is shown in Fig. 2, but now $N/2$ devices lie randomly on the circle centered at the projection of the edge server and the other $N/2$ devices lie randomly on the half circle centered at the projection of the IRS. The number of IRS reflecting elements is set to be 200. First, it can be seen that the greedy-based algorithm can only achieve a little bit better performance than that of the penalty-based algorithm. Given that the penalty-based algorithm has a very low computational complexity compared to the greedy-based algorithm, the penalty-based algorithm is more suitable for a large-scale devices system. Furthermore, it is clear that the energy consumption is increasing rapidly if we do not perform binary offloading (i.e., make all devices offload) with the aid of IRS. When the number of devices is small ($N\leq 10$), making all devices offload with the aid of IRS can perform very well. But this is not the case when the number of devices is large. This is because the communication resources of the whole system are limited (given fixed $B$ and $T$), making it difficult to support too many devices to offload even with the aid of an IRS. In essence, IRS can be regarded as a communication resource. From this perspective, the dramatic increase of the energy consumption can be postponed by increasing the number of IRS reflecting elements.
\par Fig. 6 illustrates the run times of greedy and penalty-based algorithms versus the number of devices, with the same system setup as that for generating Fig. 5.  It can be seen that when the number of devices is small ($N\leq$6), the run time of the greedy-based algorithm is comparable to that of the penalty-based algorithm. However, when the number of devices becomes relatively large, the run time of the greedy-algorithm is significantly longer than that of the penalty-based algorithm.  This corroborates the $\mathcal{O}(N)$ complexity in penalty-based algorithm versus the $\mathcal{O}(N^3)$ complexity in greedy algorithm. Thus, for a system with a small number of devices, the greedy-based algorithm is preferred. On the other hand, for a system with a large number of devices, the penalty-based algorithm achieves a quite satisfactory performance with a very short run time.
\subsubsection{Effects of IRS on different device groups}
In Fig. 7, we explore the influence to the offloading probability brought by an IRS. Under the system setup in Fig. 2, we present two groups of curves, which correspond to devices near the server and devices near the IRS, respectively. From Fig. 7, it can be observed that the penalty-based algorithm achieves a fairly close offloading probability compared with that of the greedy-based algorithm. With the increase of the number of reflecting elements, for devices near the IRS, the offloading probability is increasing monotonically to 1 with employment of an IRS because their channel conditions are significantly improved due to the proximity to the IRS. However, for devices near the server, the offloading probability decreases firstly and then remains near constant for a region and finally increases when the number of reflecting elements is large enough. The reason is that these devices are far away from the IRS, the channel conditions cannot be improved remarkably so that the offloading probability is seized by devices near the IRS when $N$ is not large enough. But when $N$ is large enough, the capability to improve the quality of wireless channel even for far away devices becomes good enough, so the offloading probability begins to increase.
\begin{figure}[htbp]
\centering
\includegraphics[width=3.5in]{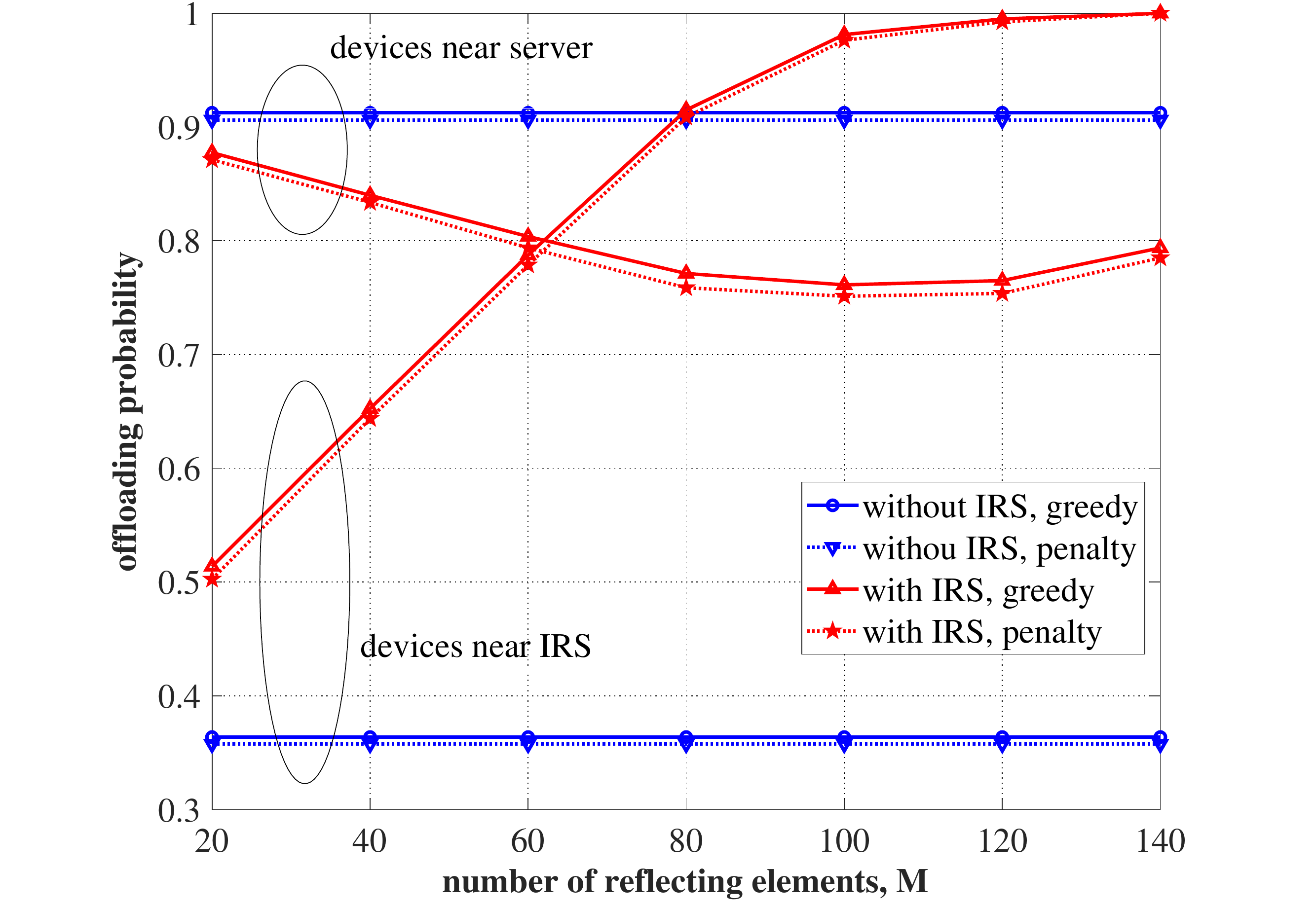}
\caption{Offloading Probability for Different Device Groups}
\end{figure}
\begin{figure}[htbp]
\centering
\includegraphics[width=3.5in]{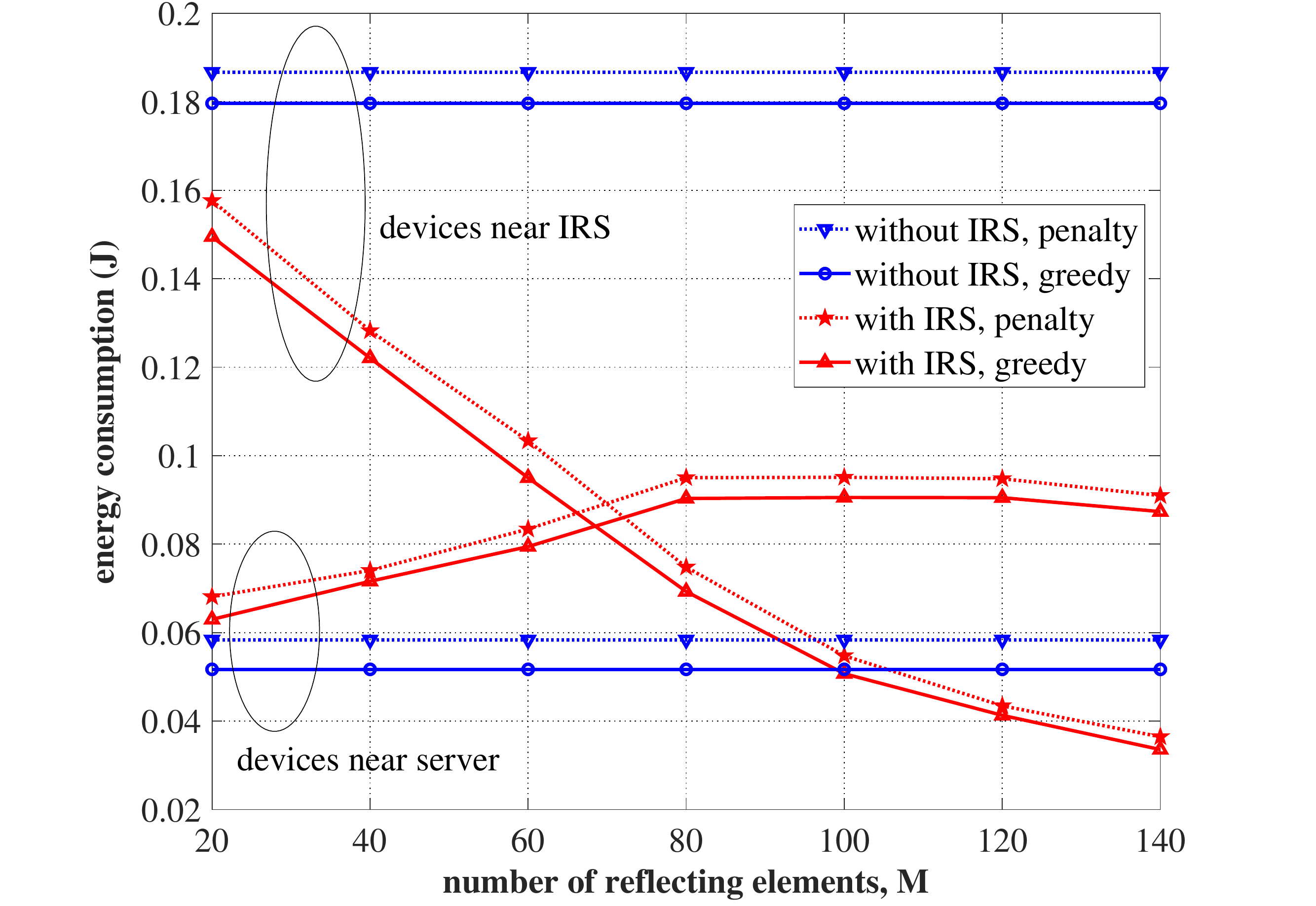}
\caption{Respective Energy Consumption for Different Device Groups}
\end{figure}
\par In Fig. 8, we illustrate the respective energy consumptions of the two device groups in Fig. 7. First, it is noticed that the penalty-based algorithm achieves a close performance to that of the greedy-based algorithm, which is consistent with the results in Fig. 3 and Fig. 5. Furthermore, with the increase of the number of reflecting elements, for devices near the IRS, it is reasonable that their energy consumption decreases monotonically since they have an increasing probability to offload the data as shown in Fig. 7. However, for devices near the server, the energy consumption increases firstly and then remains near constant and finally decreases when the number of reflecting elements is large enough. It is worth nothing that the variation trends of the energy consumptions are exactly opposite to those of the offloading probabilities in Fig. 7. From this result, it is clear that when the number of reflecting elements is not large enough, employing IRS can slightly increase the energy consumption of the devices with relatively good channel condition and significantly reduce the energy consumption of the devices with relatively bad channel condition. That is, the IRS reduces the total energy consumption of devices by balancing the energy consumption gap among devices.
\subsubsection{Impacts of discrete IRS phase shifts}
\begin{figure}[htbp]
\centering
\includegraphics[width=3.5in]{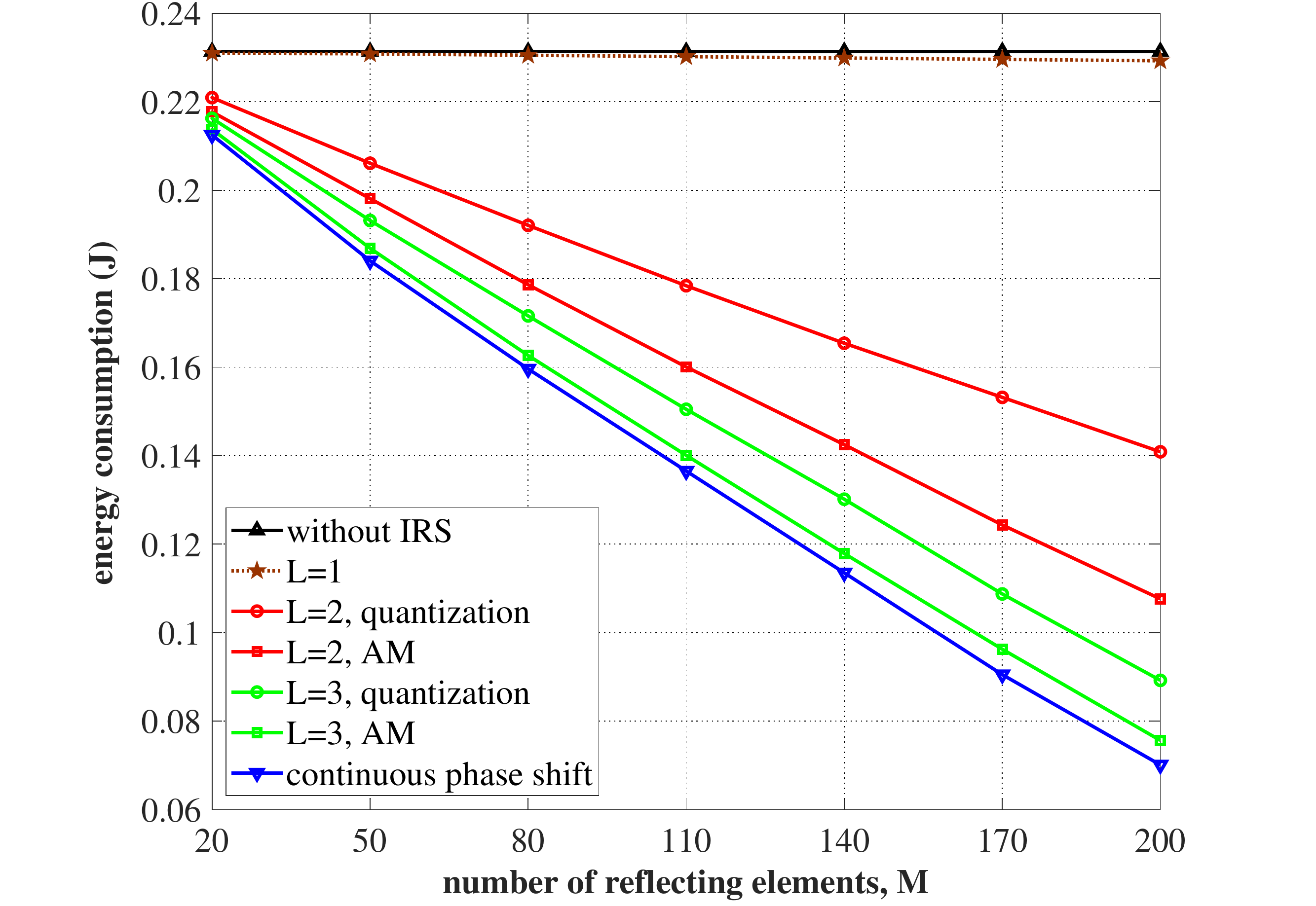}
\caption{Energy Consumption for discrete IRS phase shifts}
\end{figure}
\par Fig. 9 depicts the performance loss incurred by discrete IRS phase shifts, compared to the continuous phase shift model. The system setup is shown in Fig. 2. Without loss of generality, the greedy-based algorithm is adopted to obtain the energy consumption result. Furthermore, AM stands for the additional step of alternating maximization after quantization and quantization stands for the simple quantization method. It is clear that the additional AM step provides significant improvement over simple quantization. In particular, the AM method achieves a performance quiet close to that of continuous phase shifts even for $L=3$. Furthermore, it is noticed that the performance loss compared to continuous phase shifts increases with the increase of the number of reflecting elements. Thus, the AM method is vital to be employed to reduce the performance loss, especially for $L=2$. On the other hand, for the special scheme of $L=1$ with only one available phase shift value $0$, the IRS can only provide another signal propagation paths but cannot make the signals combine coherently at the AP. Hence, the energy consumption of $L=1$ decreases very slowly compared to other schemes when the number of reflecting elements increases.
\section{Conclusion}
This paper has proposed an IRS-aided binary offloading MEC system. An energy minimization problem has been formulated to minimize the total energy consumption of IoT devices by jointly optimizing the binary offloading modes, the CPU frequencies, the offloading powers, the offloading times and the IRS phase shifts for all devices. Greedy-based and penalty-based algorithms have been proposed to solve the challenging nonconvex problem. It was found that the greedy-based algorithm performs a little better while the penalty-based algorithm has a very low computational complexity, making these two algorithms being preferable at different scenarios. Furthermore, the deployment of IRS indeed saves more energy compared to the case without IRS.

\begin{appendices}
\section{Proof of Lemma 1}
\par For $\forall n \in \mathcal{U}_o$, define $q_n(\tau_n)=b_n \tau_n (2^{\frac {S_n} {\tau_n B}}-1)$. Then, we have $\frac{{d} q_n(\tau_n)}{{d}\tau_n}=b_n\Big(2^{\frac {S_n} {{\tau_n} B}}  (1-\ln2 \frac {S_n} {{\tau_n} B})-1\Big)$ and $\frac{{d}^2 q_n(\tau_n)}{{d}{\tau_n}^2}=(\ln2)^2 b_n 2^{\frac {2 S_n} {\tau_n B}} \frac {S_n} {{\tau_n}^2 B}$. It is obvious that $\frac{{d}^2 q_n(\tau_n)}{{d}{\tau_n}^2}>0$ when $\tau_n>0$ so $q_n(\tau_n)$ is a convex function with respect to $\tau_n>0$. Furthermore, the feasible set of $\mathcal{P}4$ is a convex set and it is a subset of $\tau_n>0$, so $q_n(\tau_n)$ is a convex function with respect to the feasible set of $\mathcal{P}4$. Thus, the objective function $\sum_{n \in \mathcal{U}_o}  b_n \tau_n (2^{\frac {S_n} {\tau_n B}}-1) + \sum_{n \in \mathcal{U}_l} \frac{\varepsilon_n S_n^3 C_n^3} {T^2}$, the summation of a set of convex functions, preserves the convexity. Therefore, $\mathcal{P}4$ is a convex optimization problem.
\par On the other hand, $\frac{{d} q_n(\tau_n)}{{d}\tau_n}=b_n\Big(2^{\frac {S_n} {{\tau_n} B}}  (1-\ln2 \frac {S_n} {{\tau_n} B})-1\Big)$ is a monotonically increasing function with respect to $\tau_n>0$ since $\frac{{d}^2 q_n(\tau_n)}{{d}{\tau_n}^2}>0$ when $\tau_n>0$. It can be observed that $\frac{{d} q_n(\tau_n)}{{d}\tau_n} \rightarrow 0$ when $\tau_n \rightarrow \infty$, thus, we assert $\frac{{d} q_n(\tau_n)}{{d}\tau_n}$ must be negative when $\tau_n>0$. Therefore, $q_n(\tau_n)$ is a monotonically decreasing function with respect to $\tau_n>0$.
\end{appendices}

\bibliographystyle{IEEEtran}
\bibliography{reference}

\begin{thebibliography}{10}
\providecommand{\url}[1]{#1}
\csname url@samestyle\endcsname
\providecommand{\newblock}{\relax}
\providecommand{\bibinfo}[2]{#2}
\providecommand{\BIBentrySTDinterwordspacing}{\spaceskip=0pt\relax}
\providecommand{\BIBentryALTinterwordstretchfactor}{4}
\providecommand{\BIBentryALTinterwordspacing}{\spaceskip=\fontdimen2\font plus
\BIBentryALTinterwordstretchfactor\fontdimen3\font minus
  \fontdimen4\font\relax}
\providecommand{\BIBforeignlanguage}[2]{{%
\expandafter\ifx\csname l@#1\endcsname\relax
\typeout{** WARNING: IEEEtran.bst: No hyphenation pattern has been}%
\typeout{** loaded for the language `#1'. Using the pattern for}%
\typeout{** the default language instead.}%
\else
\language=\csname l@#1\endcsname
\fi
#2}}
\providecommand{\BIBdecl}{\relax}
\BIBdecl

\bibitem{1}
M.~Armbrust, A.~Fox, R.Griffith, D.~Joseph, R.~Katz, A.~Konwinski, G.~Lee,
  D.~Patterson, A.Rabkin, I.Stoica, and M.Zaharia, ``A view of cloud
  computing,'' \emph{Commun. ACM}, vol.~53, no.~4, pp. 50--58, 2010.

\bibitem{2}
H.~T. Dinh, C.~Lee, D.~Niyato, and P.~Wang, ``A survey of mobile cloud
  computing: Architecture, applications, and approaches,'' \emph{Wireless
  Commun. Mobile Comput.}, vol.~13, no.~18, pp. 1587--1611, 2013.

\bibitem{31}
A.~Al-Fuqaha, M.~Guizani, M.~Mohammadi, M.~Aledhari, and M.~Ayyash, ``Internet
  of things: A survey on enabling technologies, protocols, and applications,''
  \emph{IEEE Communications Surveys \& Tutorials}, vol.~17, no.~4, pp.
  2347--2376, 2015.

\bibitem{32}
M.~R. Palattella, M.~Dohler, A.~Grieco, G.~Rizzo, J.~Torsner, T.~Engel, and
  L.~Ladid, ``Internet of things in the 5g era: Enablers, architecture, and
  business models,'' \emph{IEEE Journal on Selected Areas in Communications},
  vol.~34, no.~3, pp. 510--527, 2016.

\bibitem{33}
M.~N. Bhuiyan, D.~M.~M. Rahman, M.~M. Billah, and D.~Saha, ``Internet of things
  {(IoT)}: A review of its enabling technologies in healthcare applications,
  standards protocols, security and market opportunities,'' \emph{IEEE Internet
  of Things Journal}, 2021.

\bibitem{3}
Y.~C. Hu, M.~Patel, D.~Sabella, N.~Sprecher, and V.~Young, \emph{Mobile edge
  computing-A key technology towards 5G}.\hskip 1em plus 0.5em minus
  0.4em\relax ETSI, Sophia Antipolis, France, White Paper, 2015, vol.~11.

\bibitem{15}
W.~Shi, J.~Cao, Q.~Zhang, Y.~Li, and L.~Xu, ``Edge computing: Vision and
  challenges,'' \emph{IEEE Internet of Things Journal}, vol.~3, no.~5, pp.
  637--646, 2016.

\bibitem{16}
A.~Ahmed and E.~Ahmed, ``A survey on mobile edge computing,'' in \emph{Proc.
  IEEE ISCO}, 2016, pp. 1--8.

\bibitem{6}
P.~Mach and Z.~Becvar, ``Mobile edge computing: A survey on architecture and
  computation offloading,'' \emph{IEEE Communications Surveys \& Tutorials},
  vol.~19, no.~3, pp. 1628--1656, 2017.

\bibitem{4}
Y.~Mao, C.~You, J.~Zhang, K.~Huang, and K.~B. Letaief, ``A survey on mobile
  edge computing: The communication perspective,'' \emph{IEEE Communications
  Surveys \& Tutorials}, vol.~19, no.~4, pp. 2322--2358, 2017.

\bibitem{22}
S.~Bi and Y.~J. Zhang, ``Computation rate maximization for wireless powered
  mobile-edge computing with binary computation offloading,'' \emph{IEEE
  Transactions on Wireless Communications}, vol.~17, no.~6, pp. 4177--4190,
  2018.

\bibitem{7}
J.~Liu, Y.~Mao, J.~Zhang, and K.~B. Letaief, ``Delay-optimal computation task
  scheduling for mobile-edge computing systems,'' in \emph{Proc. IEEE Int.
  Symp. Inf. Theory (ISIT)}.\hskip 1em plus 0.5em minus 0.4em\relax Barcelona,
  Spain, 2016, pp. 1451--1455.

\bibitem{8}
Y.~Mao, J.~Zhang, and K.~B. Letaief, ``Dynamic computation offloading for
  mobile-edge computing with energy harvesting devices,'' \emph{IEEE J.Sel.
  Areas Commun.}, vol.~34, no.~12, pp. 3590--3605, 2016.

\bibitem{18}
Y.~Wang, M.~Sheng, X.~Wang, L.~Wang, and J.~Li, ``Mobile-edge computing:
  Partial computation offloading using dynamic voltage scaling,'' \emph{IEEE
  Transactions on Wireless Communications}, vol.~64, no.~10, pp. 4268--4282,
  2016.

\bibitem{21}
J.~Ren, G.~Yu, Y.~Cai, and Y.~He, ``Latency optimization for resource
  allocation in mobile-edge computation offloading,'' \emph{IEEE Transactions
  on Wireless Communications}, vol.~17, no.~8, pp. 5506--5519, 2018.

\bibitem{17}
M.~Salmani and T.~N. Davidson, ``Energy minimization of multi-user
  latency-constrained binary computation offloading,'' in \emph{Proc. IEEE
  ICASSP}, 2019, pp. 4589--4593.

\bibitem{19}
M.~Sheng, Y.~Wang, X.~Wang, and J.~Li, ``Energy-efficient multiuser partial
  computation offloading with collaboration of terminals, radio, access
  network, and edge server,'' \emph{IEEE Transactions on Communications},
  vol.~68, no.~3, pp. 1524--1537, 2020.

\bibitem{23}
C.~You, K.~Huang, H.~Chae, and B.-H. Kim, ``Energy-efficient resource
  allocation for mobile-edge computation offloading,'' \emph{IEEE Transactions
  on Wireless Communications}, vol.~13, no.~3, pp. 1397--1411, 2017.

\bibitem{5}
T.~J. Cui, M.~Q. Qi, X.~Wan, J.~Zhao, and Q.~Cheng, ``Coding metamaterials,
  digital metamaterials and programmable metamaterials,'' \emph{Light: Science
  \& Applications}, vol.~3, no.~10, p. e218, 2014.

\bibitem{9}
S.~Hu, F.~Rusek, and O.~Edfors, ``Beyond massive {MIMO}: The potential of data
  transmission with large intelligent surfaces,'' \emph{IEEE Transactions on
  Signal Processing}, vol.~66, no.~10, pp. 2746--2758, 2018.

\bibitem{10}
Q.~Wu and R.~Zhang, ``Intelligent reflecting surface enhanced wireless network:
  Joint active and passive beamforming design,'' in \emph{Proc. IEEE GLOBECOM},
  2018, pp. 1--6.

\bibitem{11}
{Q. Wu and R. Zhang}, ``Intelligent reflecting surface enhanced wireless
  network via joint active and passive beamforming,'' \emph{IEEE Transactions
  on Wireless Communications}, vol.~18, no.~11, pp. 5394--5409, 2019.

\bibitem{12}
C.~Huang, G.~C. Alexandropoulos, A.~Zappone, M.~Debbah, and C.~Yuen, ``Energy
  efficient multi-user miso communication using low resolution large
  intelligent surfaces,'' in \emph{Proc. IEEE GLOBECOM Workshops}, 2018, pp.
  1--6.

\bibitem{24}
H.~Guo, Y.~Liang, J.~Chen, and E.~G. Larsson, ``Weighted sum-rate maximization
  for reconfigurable intelligent surface aided wireless networks,'' \emph{IEEE
  Transactions on Wireless Communications}, vol.~19, no.~5, pp. 3064--3076,
  2020.

\bibitem{13}
Y.~Yang, B.~Zheng, S.~Zhang, and R.~Zhang, ``Intelligent reflecting surface
  meets {OFDM}: Protocol design and rate maximization,'' \emph{IEEE
  Transactions on Communications}, vol.~68, no.~7, pp. 4522--4535, 2020.

\bibitem{25}
B.~Zheng and R.~Zhang, ``Intelligent reflecting surface-enhanced {OFDM}:
  Channel estimation and reflection optimization,'' \emph{IEEE Wireless
  Communications Letters}, vol.~9, no.~4, pp. 518--522, 2020.

\bibitem{26}
Q.~Wu and R.~Zhang, ``Joint active and passive beamforming optimization for
  intelligent reflecting surface assisted {SWIPT} under {QoS} constraints,''
  \emph{IEEE Journal on Selected Areas in Communications}, vol.~38, no.~8, pp.
  1735--1748, 2020.

\bibitem{27}
C.~Pan, H.~Ren, K.~Wang, M.~Elkashlan, A.~Nallanathan, J.~Wang, and L.~Hanzo,
  ``Intelligent reflecting surface aided {MIMO} broadcasting for simultaneous
  wireless information and power transfer,'' \emph{IEEE Journal on Selected
  Areas in Communications}, vol.~38, no.~8, pp. 1719--1734, 2020.

\bibitem{34}
Y.~Chen, M.~Wen, E.~Basar, Y.-C. Wu, L.~Wang, and W.~Liu, ``Exploiting
  reconfigurable intelligent surfaces in edge caching: Joint hybrid beamforming
  and content placement optimization,'' \emph{IEEE Transactions on Wireless
  Communications}, 2021.

\bibitem{28}
T.~Bai, C.~Pan, Y.~Deng, M.~Elkashlan, A.~Nallanathan, and L.~Hanzo, ``Latency
  minimization for intelligent reflecting surface aided mobile edge
  computing,'' \emph{IEEE Journal on Selected Areas in Communications},
  vol.~38, no.~11, pp. 2666--2682, 2020.

\bibitem{29}
Z.~Chu, P.~Xiao, M.~Shojafar, D.~Mi, J.~Mao, and W.~Hao, ``Intelligent
  reflecting surface assisted mobile edge computing for internet of things,''
  \emph{IEEE Wireless Communications Letters}, vol.~10, no.~3, pp. 619--623,
  2021.

\bibitem{14}
L.~Z. et~al., ``Space-time-coding digital metasurfaces,'' \emph{Nat. Commun.},
  vol.~9, no.~1, p. 4334, 2018.

\bibitem{20}
S.~Boyd and L.~Vandenberghe, \emph{Convex Optimization}.\hskip 1em plus 0.5em
  minus 0.4em\relax Cambridge, U.K.: Cambridge Univ. Press, 2004.

\bibitem{30}
Q.~Wu and R.~Zhang, ``Beamforming optimization for wireless network aided by
  intelligent reflecting surface with discrete phase shifts,'' \emph{IEEE
  Transactions on Communications}, vol.~68, no.~3, pp. 1838--1851, 2020.

\end{thebibliography}
\end{document}